\documentclass[aps,prd,preprint,showpacs,superscriptaddress,groupedaddress]{revtex4-1}  
\usepackage{graphicx}  
\usepackage{float}
\usepackage{dcolumn}   
\usepackage{bm}        
\usepackage{amssymb}   
\usepackage{slashed}   
\usepackage{amsmath}   
\usepackage{simplewick} 
\usepackage{verbatim}  
\usepackage{color} 
\usepackage{appendix} 

\usepackage[bookmarks=true,colorlinks=true,linkcolor=blue,unicode=true]{hyperref}

\begin{document}

\widetext

\title{\boldmath
Time reversal symmetry violation in entangled  pseudoscalar neutral charmed mesons}
\author{Yu Shi}
\affiliation{Theoretical Physics Group, Department of Physics  \&  State Key Laboratory of Surface Physics,   Fudan University,\\ Shanghai 200433, China}
\affiliation{Collaborative Innovation Center of Advanced Microstructures, Fudan University, \\Shanghai 200433, China}
\author{Ji-Chong Yang}
\affiliation{Theoretical Physics Group, Department of Physics  \&  State Key Laboratory of Surface Physics,   Fudan University,\\ Shanghai 200433, China}
\begin{abstract}
The direct observation of time reversal symmetry violation (TV) is important for the test of $CPT$ conservation and the Standard Model. In  this paper, we study both time-dependent and time-independent genuine TV signals in entangled $D^0-\bar{D}^0$ pairs. A possible $CPT$-violation effect called the $\omega$ effect is also investigated. In the $C=-1$ entangled state,  the asymmetries due to TV  are  calculated to be  of the order of $10^{-5}$ to $10^{-4}$  within the Standard Model, but the modification due to the $\omega$ effect in the $C=-1$  states is found to be  about $10\%-30\%$ when $|\omega|\sim 10^{-4}$. This result is consistent with our Monte Carlo simulation, which implies that with $10^9$  to   $10^{10}$ events, TV signals can be observed in the entangled $D^0-\bar{D}^0$ pairs, and  the bound of $\omega \sim 10^{-3}$ can be reached.  The time-dependent and the time-independent  asymmetries in the $C=-1$ $D^0-\bar{D}^0$ system  provides a window to detect new physics such as the $\omega$ effect, although they are not easily observable.
\end{abstract}


\maketitle

\section{\label{sec:level1}Introduction}

Symmetry, symmetry violation and symmetry breaking have been playing important roles in particle physics. The studies of discrete symmetries $P$, $C$, $T$ and their combinations have progressed greatly with the help of large experimental data~\cite{CPreview}. There are oscillations between neutral mesons  and their antiparticles, such as $B^0-\bar{B}^0$, $D^0-\bar{D}^0$ and $K^0-\bar{K}^0$. In the $D^0-\bar{D}^0$ system, both the mass and the decay width differences between the two mass eigenstates are very small in comparison with the mean values~\cite{PDG}. This provides an opportunity to verify $CP$ violation (CPV) sources from both the Standard Model (SM) and new physics (NP)~\cite{NP}  and even  the possibility of  $CPT$ violation (CPTV)  such as the so-called $\omega$ effect, as predicted by some theories of quantum gravity~\cite{CPTV,omega}.

If $CPT$ is conserved~\cite{CPT}, then CPV implies time reversal (T) symmetry violation (TV). However, direct observation of TV without the presumption of $CPT$ conservation is especially important~\cite{original,Treview1,kextension,Treview2}. The TV signal based on a T-odd product of momentum vectors was observed in the decay $D^0 \to K^+K^-\pi^+\pi^-$~\cite{TV}. However, such a signal has a chance of being nongenuine because the initial and final states are not interchanged~\cite{Treview1}. The TV signal based on the rate difference between the transformation from $K^0$ to $\bar{K}^0$ and vice versa~\cite{TVK} is controversial~\cite{Treview1}.

Hence an important development is that a genuine TV signal has been observed in $B^0-\bar{B}^0$ decay, by comparing transitions that are related through time reversal but not through $CP$ conjugation~\cite{TVB,TVBDiscuss}. The key idea is to make use of quantum entanglement, also called the Einstein-Podolsky-Rosen correlation~\cite{original,Treview1,kextension}. The initial states of each of the two transitions  is prepared by tagging the entangled partners in the corresponding way.  The connections between $CP$, $T$ and $CPT$ asymmetries and the experimental asymmetries  are  investigated for entangled $B_d^0\bar{B}_d^0$ mesons~\cite{BdSystem}. Extension to kaons has  been made~\cite{kextension,k2}.

In this paper, we propose using the time-independent signals to study TV by extending the entanglement approach of TV to $D^0-\bar{D}^0$ systems. The $C=-1$ entangled $D^0-\bar{D}^0$ pairs can be produced through the strong decay of $\psi (3770)$~\cite{BESCII,JDRxzz1,JDRxzz2} or $\psi (4140)$~\cite{JDRxzz1,JDRxzz2}.  $\psi (3770)$ has often been used for the study of CPV of $D$ mesons. The $C=+1$ entangled state of $D$ mesons can also be produced in the strong decay of $\psi(4140)$~\cite{JDRxzz1,JDRxzz2}.

First, we calculate the time-dependent and the time-independent asymmetries between T-conjugate  processes for the $C=-1$ entangled  states. Within the SM, the asymmetry of the $C=-1$ system is found to be at most $10^{-5}$. We also consider the $\omega$ effect  in the $C=-1$ state, which mixes the $C=+1$ state into it.  We find that the  $\omega$ effect modifies the TV signals by as large as $20\%$ when $|\omega|\sim 10^{-4}$. We also calculate the T asymmetries defined for transitions from $D^0$ to $D^-$ and vice versa,  by using event numbers in joint decays of entangled pairs. Finally we use a Monte Carlo simulation~\cite{mcmethod} to study the $C=-1$ systems based on the current experimental situation, and demonstrate that if the number of events reachea $10^9$ TV signals can be observed; furthermore,  if the number of events reaches $10^{10}$, the bound of $\omega \sim 10^{-3}$ can be obtained.

We conclude that in the $C=-1$ $D^0-\bar{D}^0$ entangled state the time-dependent asymmetry due to TV within the SM requires a large number of events and may provide a window to detect the signal of NP such as the $\omega$ effect.

The rest of this paper is organized as follows. In Sec.~\ref{sec:level2} we briefly review the idea of studying TV using the  entangled states.  In Sec.~\ref{sec:level3}, we study the joint decay rates of such states.  In Sec.~\ref{sec:level4}, we discuss the  TV signals  in the oscillation of the $D^0-\bar{D}^0$ system. Section.~\ref{sec:level5} is a discussion on the relation between the joint decay rate and the experimental measurement. In Sec.~\ref{sec:level6}, we present a Monte Carlo simulation on  the TV.  Section.~\ref{sec:level7} is a summary.

\section{\label{sec:level2} Entangled states of neutral mesons  }

As pseudoscalar neutral mesons consisting of quarks, $D^0=c\bar{u}$ and $\bar{D}^0 = \bar{c}u$. In the Wigner-Weisskopf approximation,  $|D^0\rangle$ and $|\bar{D}^0\rangle$ are eigenstates of the flavor, which is the charm in this specific case, with eigenvalues $\pm 1$. $|D^0\rangle$ and $|\bar{D}^0\rangle$ comprise  a  basis, in which  the  effective mass matrix is written as
 \begin{equation}
H=\left(
\begin{array}{cc}
H_{00} & H_{0\bar{0}} \\
H_{\bar{0}0} & H_{\bar{0}\bar{0}}
\end{array}\right),
\end{equation}
where $H_{00}\equiv \langle D^0|H|D^0\rangle$, $H_{0\bar{0}}\equiv \langle D^0|H|\bar{D}^0\rangle$, and so on. The eigenstates of $H$  are
\begin{equation}
\begin{split}
&|D_H\rangle=p|D^0\rangle + q|\bar{D}^0\rangle,\;\;\;|D_L\rangle=p|D^0\rangle - q|\bar{D}^0\rangle,
\end{split}
\label{eq.3.1}
\end{equation}
with
\begin{equation}
\frac{p}{q} =\frac{1-\epsilon}{1+\epsilon}=
\sqrt{ \frac{H_{\bar{0}0}}{ H_{0\bar{0}}} },
\end{equation}
where $\epsilon$ is the indirect CPV parameter. The corresponding
eigenvalues are
\begin{equation}
\begin{split}
\lambda _{H}&= m_{H}-\frac{i}{2}\Gamma _{H}=H_{00} +\sqrt{ H_{0\bar{0}}H_{\bar{0}0} },\\
 \lambda _{L}&= m_{L}-\frac{i}{2}\Gamma _{L}
 =H_{00} - \sqrt{ H_{0\bar{0}}H_{\bar{0}0} },
\end{split}
\label{eq.3.4}
\end{equation}

We can neglect the direct CPV, as done in  testing T violation in entangled $B$ mesons~\cite{mcmethod,Treview1,BdSystem}, and can also be done in  entangled $D$ mesons~\cite{PDG,Af,BESCII}.

We will use the definitions
\begin{equation}
\begin{split}
&\Delta m \equiv m_H-m_L,\;\;\;   \Delta \lambda \equiv \lambda _H-\lambda _L,\;\;\;  \Delta \Gamma \equiv \Gamma_H-\Gamma_L,\\
&m \equiv \frac{1}{2}( m_H+m_L),\;\;\;\Gamma \equiv \frac{1}{2}(\Gamma_L+\Gamma_H). \\
\end{split}
\label{eq.3.gammamass}
\end{equation}
The sign of $\Delta \Gamma$ in the definition (\ref{eq.3.gammamass}) is different from $\Delta \Gamma$ defined in Refs.~\cite{JDRxzz1,JDRreview} and is same as in Refs.~\cite{PDG,JDRyang,cplus,HFAG}.

The time evolution of  the mass eigenstates is
\begin{equation}
\begin{split} &|D_H(t)\rangle\equiv U(t)|D_H\rangle=e^{-i\lambda_H t} |D_H\rangle,\;\;
|D_L(t)\rangle\equiv U(t)|D_L\rangle=e^{-i\lambda_L t} |D_L\rangle,
\end{split}
\end{equation}
where $U(t)$ represents the time evolution under the effective mass matrix. $U(t)$ evolves the  flavor basis states as
\begin{equation}
\begin{split}
&|D^0(t)\rangle\equiv U(t)|D^0\rangle= g_+(t)|D^0\rangle
-\frac{q}{p}g_-(t)|\bar{D}^0\rangle,\\
&|\bar{D}^0(t)\rangle\equiv U(t)|\bar{D}^0\rangle=
-\frac{p}{q}g_-(t)|D^0\rangle+g_+(t)|\bar{D}^0\rangle,\\
\end{split}
\label{eq.3.2}
\end{equation}
with
\begin{equation}
g_{\pm}(t) \equiv \frac{e^{-i\lambda _Lt} \pm e^{-i\lambda _Ht}}{2},  \label{eq.3.3}
\end{equation}
where the sign of $g_-(t)$  is different from that  in Ref.~\cite{JDRyang}, and is same as in Refs.~\cite{JDRxzz1,JDRreview}. The more general expressions of $|D^0(t)\rangle$ and $|\bar{D}^0(t)\rangle$, without the assumption of indirect $CPT$ conservation,  are   given in Refs.~\cite{JDRshi1,cplus}, and reduce to the expressions here when $CPT$ is indirectly conserved.
Note that the two mass eigenstates $|D_H\rangle$ and $|D_L\rangle$ are not orthogonal because of indirect CPV parameter $\epsilon \neq 0$; hence, the basis transformation involving them is not unitary.

There is yet another basis often used, namely, the $CP$ basis,
\begin{equation}
|D_{\pm}\rangle = \frac{1}{\sqrt{2}}\left(|D^0\rangle \pm |\bar{D}^0\rangle\right).
\label{eq.1.3}
\end{equation}
with eigenvalue $\pm 1$. The   time evolution starting with each of them can be written as
\begin{equation}
|D_{\pm}(t)\rangle = U(t)|D_{\pm}\rangle.
\end{equation}

Now suppose at time $t=0$, the $C=\pm 1$ entangled states of two mesons $a$ and $b$ is generated,
\begin{equation}
|\Psi _C\rangle = \frac{1}{\sqrt{2}}\left(|D^0\rangle_a |\bar{D}^0\rangle_b + C |\bar{D}^0\rangle_a|D^0\rangle_b \right).
\label{eq.3.5}
\end{equation}
where the subscripts $a$ and $b$ will be omitted below. Under the mass matrix, $|\Psi _C\rangle $  evolves to
\begin{equation}
|\Psi _C(t)\rangle = \frac{1}{\sqrt{2}}\left(|D^0(t)\rangle_a |\bar{D}^0(t)\rangle_b + C |\bar{D}^0(t)\rangle_a|D^0(t)\rangle_b \right).
\end{equation}
Specifically£¬
\begin{equation}
|\Psi_-(t)\rangle = e^{-i(\lambda_H+\lambda_L)t}|\Psi_-\rangle,
\end{equation}
\begin{equation}
\begin{split}
&|\Psi _+(t)\rangle=\frac{e^{-i(\lambda _H+\lambda _L)t}}{\frac{2q}{p}}\left(\frac{q}{p}\cos(\Delta \lambda  t)( |D^0\rangle |\bar{D}^0\rangle \right.\\
&\left. +  |\bar{D}^0\rangle |D^0\rangle) + i\sin(\Delta \lambda t) (|D^0\rangle |D^0\rangle +\left(\frac{q}{p}\right)^2 |\bar{D}^0\rangle |\bar{D}^0\rangle )\right). \\
\end{split}
\label{plus}
\end{equation}
It can be seen that the evolution of $\Psi _-$ leaves the entanglement unchanged and provides a good opportunity to study the discrete symmetries. Furthermore, one can define
\begin{equation}
\begin{split}
&|\Psi_C(t_a,t_b) \rangle \equiv  U(t_b)U(t_a)|\Psi_C\rangle =\frac{1}{\sqrt{2}}\left(|D^0(t_a)\rangle |\bar{D}^0(t_b)\rangle+C|\bar{D}^0(t_a)\rangle|D^0(t_b)\rangle \right),
\end{split}
\label{eq.1.1}
\end{equation}
which represents that particle a decays at $t_a$ while particle b decays at $t_b$, and is widely used in calculating joint decay rate~\cite{Treview1,entangled,JDRxzz1}.  $|\Psi_C(t_a,t_b)$ can also be written in terms of $CP$ eigenstates as
\begin{equation}
\begin{split}
&|\Psi_-(t_a,t_b) \rangle =  \frac{1}{\sqrt{2}}\left(|D_-(t_a)\rangle|D_+(t_b)\rangle -|D_+(t_a)\rangle|D_-(t_b)\rangle \right),\\
&|\Psi_+(t_a,t_b) \rangle =  \frac{1}{\sqrt{2}}\left(|D_+(t_a)\rangle|D_+(t_b)\rangle -|D_-(t_a)\rangle|D_-(t_b)\rangle \right).
\end{split}
\label{eq.1.2}
\end{equation}

Unless explicitly stated,  here, $t_b\geq t_a$ is assumed without loss of generality. The free choice between Eqs.~(\ref{eq.1.1}) and (\ref{eq.1.2}) can be made by determining whether the earlier decay of  meson  $a$  is into a $CP$ eigenstate or a flavor eigenstate. We use $l^{\pm}$ to denote a final state of a semileptonic decay with flavor number $\pm 1$ and $S_\pm$ to denote the final state of a $CP$ eigenstate with eigenvalue $\pm 1$.

\section{\label{sec:level3}  T-conjugate transitions obtained from the entangled mesons }

\subsection{\label{sec:300} T-conjugate transitions}

The entangled meson pairs can be used in  the so-called single-tag~(ST) and double-tag~(DT) methods~\cite{DT,DT2,BESCII}. In the case of the $C=-1$ entangled state, the final state of the first  decay at $t_a$ tags the partner as $D^0$ or $\bar{D}^0$ or $D_{\pm}$; one can then study the decay  of the tagged partner  at $t_b$.

Because  at time $t$, $|\Psi_-(t)\rangle \propto |\Psi_-\rangle$, the $C=-1$ entangled state can be used to construct T-conjugate processes.
For example, if meson $a$ decays into the $l^-$ final state at $t_a$, it implies that meson $a$ has been projected to $|\bar{D}^0\rangle$, which decays to the $l^-$ final state; hence, meson $b$ is prepared to be $|D^0\rangle$ at $t_a$. Then, by measuring the probability that meson $b$ decays into a final state $S_-$ at a later time $t_b$, one obtains the probability that meson $b$ evolves and then transits  to $D_-$ during the time period $t_b-t_a$.

Therefore the final states of the two entangled mesons act as tags. With the help of the tags, one can measure the rate of the transition $D^0  \to D_- $ of  meson $b$.

The time reversal symmetry requires that the transition rate of $D^0 \to D_-$ from $t$ to $t+\Delta t$ is equal to that of $D_- \to D^0$ from $t'$ to $t'+\Delta t$. They can be prepared  alternatively as the transitions of meson $b$ through double tags. For the process $D^0 \to D_-$,  $D^0$, as the initial state of meson $b$, is prepared when  the final state of meson $a$ is $l^-$, while $D_-$ is indicated by the final state $S^-$ of meson $b$ (with the direct CPV neglected). For the process $D_- \to D^0$,   $D_-$, as the initial state of meson $b$, is prepared when the final state of meson $a$  is $S^+$, while  $D^0$ is indicated by the final state $l^+$ of meson $b$. Various transitions and the corresponding final states, as used to observe the TV, are summarized in Table~\ref{tab1}~\cite{Treview1,mcmethod}.

To test TV, we need to compare these T-conjugation transitions. There are several ways to relate the transitions to observables, as discussed below.

\begin{table}
\caption{$T$-conjugate transitions and the corresponding final states of the $C=-1$ entangled mesons $a$ and $b$. We use $l^{\pm}$ to denote a final state of a semileptonic decay, with flavor number $\pm 1$, and use $S_{\pm}$ to denote the $CP$ eigenstate with eigenvalue $\pm 1$.   Meson $a$ decays at $t_a$, while meson $b$ decays at a later time  $t_b \geq t_a$. The transition listed is that of  meson $b$.}
\label{tab1}

\begin{tabular}{  p{1.9cm} | p{3.5cm} | p{1.9cm} | | p{1.9cm} |  p{3.5cm} | p{1.9cm} }
\hline\noalign{\smallskip}
Final state of meson $a$ &  Transition of meson $b$ & Final state of meson $b$  & Final state of meson $a$  & T-conjugate transition of meson $b$ & Final state of meson $b$ \\
\noalign{\smallskip}\hline\noalign{\smallskip}
$l^-$ &$D^0\to D_-$ &  $S_-$ &  $S_+$ & $D_-\to D^0$ & $l^+$ \\  $l^-$ &
$D^0\to D_+$ & $S_+$ & $S_-$ &  $D_+\to D^0$ &  $l^+$ \\ $l^+$ &
$\bar{D}^0\to D_-$ &  $S_-$ & $S_+$ &  $D_-\to \bar{D}^0$ & $l^-$ \\ $l^+$ &
$\bar{D}^0\to D_+$ & $S_+$ & $S_-$ &  $D_+\to \bar{D}^0$ &  $l^-$ \\
\hline
\noalign{\smallskip}\hline
\end{tabular}
\end{table}

\subsection{\label{sec:level30}Joint decay rates}

For  the entangled meson pairs, an  important quantity to study is the joint decay rate, which  is the joint rate of the  processes  in which one of the entangled mesons decays into the final state $f_a$ at $t_a$ while the other decays into $f_b$ at $t_b$~\cite{JDRshi1,JDRyang,JDRxzz1,JDRreview,cplus}. The   rate $\Gamma_C(f _a,f _b, t_a, t_b)$ at which meson $a$ decays to $f_a$ at $t_a$ while $b$  decays to $f_b$ at $t_b$ is proportional to the joint decay rate calculated from $|\Psi _C(t_a,t_b)\rangle$,
\begin{equation}
\begin{split}
&\Gamma_C(f _a,f _b, t_a, t_b)\propto R_C(f _a,f _b, t_a, t_b)\equiv \left|\langle f _a, f _b|{\cal H}_a{\cal H}_b|\Psi _C(t_a,t_b)\rangle \right|^2.
\end{split}
\end{equation}

The rate of each transition listed in Table~\ref{tab1} can be obtained from the joint decay rate of the corresponding final states, with   meson $a$ decaying to its final state such that the entangled partner $b$ is projected to the initial state in the transition listed.

\subsubsection{\label{sec3.1} Joint decay rates of \texorpdfstring{$C=\pm 1$}{C=+-1} states}

For $|\Psi _C(t_a,t_b)\rangle$,  the joint decay amplitude for the joint processes in which meson $a$ decays to $f_a$ at $t_a$ while meson $b$ decays to $f_b$  at $t_b$ is
\begin{equation}
\begin{split}
&\langle f _a, f _b|{\cal H}_a{\cal H}_b|\Psi _C(t_a,t_b)\rangle \\ &=\frac{1}{\sqrt{2}}\left\{\xi_C\left[g_+(t_a)g_-(t_b)+Cg_-(t_a)g_+(t_b)\right]+\zeta_C\left[g_+(t_a)g_+(t_b)+Cg_-(t_a)g_-(t_b)\right]\right\},
\end{split}
\label{eq.3.6}
\end{equation}
where ${\cal H}_a$ is the weak interaction field theoretic   Hamiltonian governing the decay of the meson $a$ and $\xi _C$ and $\zeta_C$ are defined as
\begin{equation}
\begin{split}
&\xi_C\equiv-\left(\frac{p}{q}A_{f_a}A_{f_b}
+C\frac{q}{p}\bar{A}_{f_a}\bar{A}_{f_b}\right),\;\;\zeta_C\equiv A_{f_a}\bar{A}_{f_b}+C\bar{A}_{f_a}A_{f_b},\\
\end{split}
\label{eq.3.7}
\end{equation}
where $A_f$ and $\bar{A}_f$ are instantaneous decay amplitudes
\begin{equation}
\begin{split}
&A_f\equiv\langle f |{\cal H}|  D^0 \rangle,\;\; \bar{A}_f\equiv\langle f |{\cal H} |\bar{D}^0\rangle .
\end{split}
\label{eq.3.8}
\end{equation}

The joint decay rate is thus
\begin{equation}
\begin{array}{rll}
&R_C(f _a,f _b, t_a, t_b)=\left|\langle f _a, f _b|{\cal H}_a{\cal H}_b|\Psi _C(t_a,t_b)\rangle \right|^2 \\
&=\frac{e^{-\Gamma (t_a+t_b)}}{4}\times \left\{(|\xi_C|^2+|\zeta_C|^2)\cosh (y\Gamma (t_a+Ct_b))-(|\xi_C|^2-|\zeta_C|^2)\cos (x\Gamma (t_a+Ct_b))\right.\\
&\left.+2CRe(\zeta_C^*\xi_C)\sinh (y\Gamma (t_a+Ct_b))-2CIm(\zeta_C^*\xi_C)\sin (x \Gamma (t_a+Ct_b))\right\},\\
\end{array}
\label{eq.3.9}
\end{equation}
where $x$ and $y$ are defined as
\begin{equation}
\begin{split}
&x\equiv\frac{\Delta m}{\Gamma},\;\;y\equiv \frac{\Delta \Gamma}{2\Gamma}.
\end{split}
\label{eq.3.11}
\end{equation}

In experiments, we often use the time-integrated joint decay
\begin{equation}
\begin{split}
&R_C(f _a,f _b,\Delta t)=\int _0^{\infty}dt_aR_C(f _a,f _b, t_a, t_a+\Delta t)\\
\end{split}
\label{eq.3.12}
\end{equation}
hence
\begin{equation}
\begin{split}
&R_C(f _a,f _b,\Delta t >0)= \frac{e^{-\Gamma  \Delta t }}{8\Gamma}\\
&\times \left\{(|\xi_C|^2+|\zeta_C|^2)\left[\cosh (y\Gamma \Delta t)+\left(\frac{1+C}{2}\right)\frac{y^2\cosh (y\Gamma \Delta t)+y \sinh(y\Gamma \Delta t)}{1-y^2}\right] \right.\\
&\left.- (|\xi_C|^2-|\zeta_C|^2)\left[\cos (x\Gamma \Delta t)+\left(\frac{1+C}{2}\right)\frac{-x^2\cos (x\Gamma \Delta t)-x\sin (x\Gamma \Delta t)}{1+x^2}\right]\right.\\
&\left.+2{\rm Re}(\zeta_C^*\xi_C)\left[\sinh(y\Gamma \Delta t)+\left(\frac{1+C}{2}\right)\frac{y \cosh (y\Gamma \Delta t)+y^2 \sinh(y\Gamma \Delta t)}{1-y^2}\right]\right.\\
&\left.-2{\rm Im}(\zeta_C^*\xi_C)\left[\sin (x\Gamma \Delta t)+\left(\frac{1+C}{2}\right)\frac{x\cos (x\Gamma \Delta t)-x^2\sin (x\Gamma \Delta t)}{1+x^2}\right]\right\}.\\
\end{split}
\label{eq.3.13}
\end{equation}

Finally, the time-independent joint decay rate is defined as
\begin{equation}
\begin{split}
&R_C(f _a,f _b) \equiv \int _0^{\infty}dt_a\int _0^{\infty}dt_b\left|\langle f _a, f _b|{\cal H}_a{\cal H}_b|\Psi _C(t_a,t_b)\rangle \right|^2,\\
\end{split}
\label{eq.3.14}
\end{equation}
which is obtained as
\begin{equation}
\begin{split}
&R_C(f _a,f _b)= \frac{1}{4\Gamma^2}\left((|\xi_C|^2+|\zeta_C|^2)\frac{1+Cy^2}{(1-y^2)^2}-(|\xi_C|^2-|\zeta_C|^2)\frac{1-Cx^2}{(1+x^2)^2}\right.\\
&\left.+2Re(\zeta_C^*\xi_C)\frac{(1+C)y}{(1-y^2)^2}-2Im(\zeta_C^*\xi_C)\frac{(1+C)x}{(1+x^2)^2}\right).\\
\end{split}
\label{eq.3.15}
\end{equation}
Note that $R_C(f _a,f _b)$ is independent of the order of the two final states. In experiments, such time-independent quantities are most easily measured.

\subsubsection{\label{sec3.2}Joint decay rates under the \texorpdfstring{$\omega$ effect}{omega effect}.}

One kind of CPTV is the so-called $\omega$ effect, which is a consequence of some forms of quantum gravity~\cite{CPTV,omega}. The $\omega$ effect affects the entangled source, so the $C=-1$ entangled state is mixed in by  the $C=+1$ entangled state with a factor $\omega$. For simplicity, in this section we assume the CPV parameters are barely affected by the $\omega$ effect.

Because of the $\omega$ effect, the $C=-1$ entangled state is  modified to be
\begin{equation}
\begin{split}
&|\Psi _{\omega}(t_a,t_b)\rangle = |\Psi _-(t_a,t_b)\rangle+\omega |\Psi _+(t_a,t_b)\rangle ,\\
\end{split}
\label{eq.3.16}
\end{equation}
where $$\omega \equiv |\omega|e^{i\Omega}$$ is a small mixing factor. The joint decay rate is found to be
\begin{equation}
\begin{split}
&R_{\omega}(f _a,f _b, t_a, t_b)=R_-(f _a,f _b, t_a, t_b)+|\omega|^2R_+(f _a,f _b, t_a, t_b)+R_m(f _a,f _b, t_a, t_b),
\end{split}
\label{eq.3.17}
\end{equation}
with
\begin{equation}
\begin{split}
&R_m(f _a,f _b, t_a, t_b)\\
& \equiv e^{-\Gamma (t_a+t_b)} \left[
  {\rm Re}(\alpha+\beta)\cos (x\Gamma t_a)\cosh(y\Gamma t_b)
  -{\rm Im}(\alpha+\beta)\sin (x\Gamma t_a)\sinh(y\Gamma t_b)\right.\\
&\left.
 -{\rm Re}(\alpha-\beta)\cos (x\Gamma t_b)\cosh(y\Gamma t_a)
  +{\rm Im}(\alpha-\beta)\sin (x\Gamma t_b)\sinh(y\Gamma t_a)\right.\\
&\left.
  +{\rm Re}(\rho+\sigma)\cos(x\Gamma t_a)\sinh (y\Gamma t_b)
  -{\rm Im}(\rho+\sigma)\sin(x\Gamma t_a)\cosh (y\Gamma t_b)\right.\\
&\left.
 -{\rm Re}(\rho-\sigma)\cos(x\Gamma t_b)\sinh (y\Gamma t_a)
  +{\rm Im}(\rho-\sigma)\sin(x\Gamma t_b)\cosh (y\Gamma t_a)\right],\\
\end{split}
\label{eq.3.18}
\end{equation}
where
\begin{equation}
\begin{split}
&\alpha\equiv\frac{\omega}{2}\xi_-^*\xi_+,\;\;\;
 \beta\equiv\frac{\omega}{2}\zeta_-^*\zeta_+,\;\;\;
 \rho\equiv\frac{\omega}{2}\xi_-^*\zeta_+,\;\;\;
 \sigma\equiv\frac{\omega}{2}\zeta_-^*\xi_+.\\
\end{split}
\label{eq.3.19}
\end{equation}
The integrated joint decay rate can be written as
\begin{equation}
\begin{split}
&R_{\omega}(f _a,f _b, \Delta t)=R_-(f _a,f _b, \Delta t)+|\omega|^2R_+(f _a,f _b, \Delta t)+R_m(f _a,f _b, \Delta t),
\end{split}
\label{eq.3.20}
\end{equation}
where
\begin{equation}
\begin{split}
&R_m(f_a,f_b,\Delta t)=\frac{e^{-\Gamma\Delta t}}{\Gamma}\left({\rm Ch}\cosh(y\Gamma \Delta t)+{\rm Sh}\sinh(y\Gamma \Delta t)+{\rm Cs}\cos(x \Gamma \Delta t)+{\rm Sn}\sin(x\Gamma \Delta t)\right),
\end{split}
\label{eq.3.21}
\end{equation}
with
\begin{equation}
\begin{split}
&{\rm Ch}\equiv A{\rm Re}(\alpha+\beta)-D{\rm Im}(\alpha+\beta)-yB{\rm Re}(\rho+\sigma)-xC{\rm Im}(\rho+\sigma),\\
&{\rm Sh}\equiv -yB{\rm Re}(\alpha+\beta)-xC{\rm Im}(\alpha+\beta)+A{\rm Re}(\rho+\sigma)-D{\rm Im}(\rho+\sigma),\\
&{\rm Cs}\equiv -A{\rm Re}(\alpha-\beta)+D{\rm Im}(\alpha-\beta)+yB{\rm Re}(\rho-\sigma)+xC{\rm Im}(\rho-\sigma),\\
&{\rm Sn}\equiv xC{\rm Re}(\alpha-\beta)-yB{\rm Im}(\alpha-\beta)+D{\rm Re}(\rho-\sigma)+A{\rm Im}(\rho-\sigma),\\
&A\equiv\frac{2(x^2-y^2+4)}{x^4+2 x^2 \left(y^2+4\right)+\left(y^2-4\right)^2},
 B\equiv\frac{\left(x^2+y^2-4\right)}{x^4+2 x^2 \left(y^2+4\right)+\left(y^2-4\right)^2},\\
&C\equiv\frac{\left(x^2+y^2+4\right)}{x^4+2 x^2 \left(y^2+4\right)+\left(y^2-4\right)^2},
 D\equiv\frac{4x y}{x^4+2 x^2 \left(y^2+4\right)+\left(y^2-4\right)^2}.
\end{split}
\label{eq.3.22}
\end{equation}

\section{\label{sec:level4}TV signals in \texorpdfstring{$D^0-\bar{D}^0$}{D0 D0 bar} systems  }

In this section, we first establish the TV signals and their behavior  predicted within the SM. We use those decay channels in which the direct CPV, i.e. that in the decays, can be neglected and only consider indirect CPV, i.e., that in the oscillation. We consider only the cases in which one of the final states is a $CP$ eigenstate while the other is a flavor eigenstate~\cite{Treview1,mcmethod,BESCII}. In $D^0-\bar{D}^0$ systems, the indirect CPV parameter is known to be very small~\cite{PDG,CPVSummery}. Within the SM, the corresponding TV is also expected to be very small.

With direct CPV negligible, we have~\cite{PDG,Af}
\begin{equation}
\begin{split}
&A_{l^-}=\bar{A}_{l^+}=0,\;\;\;A_{l^+}=\bar{A}_{l^-}\equiv A_l.\\
\end{split}
\label{eq.4.1}
\end{equation}

When the final state is a $CP$ eigenstate $S_{\pm}$, within the SM, we have~\cite{Af}
\begin{equation}
\begin{split}
&A_{S_{\pm}}=\pm \bar{A}_{S_{\pm}},
\end{split}
\label{eq.4.2}
\end{equation}

where $A_f$ and $\bar{A}_f$ are defined in Eq.~(\ref{eq.3.9}). Substituting $(f_a,f_b)=(l^{\pm},S_{\pm})$   in Eq.~(\ref{eq.3.7}), we find
\begin{equation}
\begin{split}
&|\xi_C|^2+|\zeta_C|^2=|A_l|^2 |A_{S_{\pm}}|^2\left(\left|\frac{p}{q}\right|^{2n_l}+1\right),\\
&|\xi_C|^2-|\zeta_C|^2=|A_l|^2 |A_{S_{\pm}}|^2\left(\left|\frac{p}{q}\right|^{2n_l}-1\right),\\
&2{\rm Re}(\zeta_C^*\xi_C)= -2n_s\left|\frac{p}{q}\right|^{n_l} \cos (2\phi) |A_l|^2 |A_{S_{\pm}}|^2,\\
&2{\rm Im}(\zeta_C^*\xi_C)= 2n_ln_s\left|\frac{p}{q}\right|^{n_l} \sin (2\phi) |A_l|^2 |A_{S_{\pm}}|^2,
\end{split}
\label{eq.4.4}
\end{equation}
where $n_l=\pm 1$ for $l^{\pm}$ final states and $n_s=\pm 1$ for $S_{\pm}$ states.

Experimentally, the semileptonic decay modes and the $CP$ eigenstate decay modes of a $C=-1$ entangled $D^0-\bar{D}^0$ system have been studied by using DT of the two mesons~\cite{BESCII}, where the semileptonic decay modes include $Ke\nu$ and $K\mu\nu$, while the $CP$ eigenstate decay modes include  $K^+K^-$, $\pi ^+\pi ^-$, and  $K^0_S\pi^0\pi^0$   for $CP=1$ and  $K^0_S\pi^0$, $K^0_S\omega$, and $K^0_S\eta$ for $CP=-1$.

$q/p$ is often parametrized as
\begin{equation}
\begin{split}
&\frac{q}{p}\equiv \left|\frac{q}{p}\right|e^{i2\phi},\\
\end{split}
\label{eq.4.3}
\end{equation}
which will be used below. Other frequently used parameters include $y_{\rm CP}$ and $A_{\Gamma}$ which can be defined as~\cite{PDG,BESCII,YcpDefine}
\begin{equation}
\begin{split}
&y_{\rm CP}\equiv \frac{1}{2}\left(y\cos (2\phi) \left(\left|\frac{q}{p}\right|+\left|\frac{p}{q}\right|\right)-x\sin (2\phi) \left(\left|\frac{q}{p}\right|-\left|\frac{p}{q}\right|\right)\right),\\
&A_{\Gamma}\equiv \frac{1}{2}\left(y\cos (2\phi) \left(\left|\frac{q}{p}\right|-\left|\frac{p}{q}\right|\right)-x\sin (2\phi) \left(\left|\frac{q}{p}\right|+\left|\frac{p}{q}\right|\right)\right).\\
\end{split}
\label{eq.4.ycpdefine}
\end{equation}
$y_{CP}\neq y$ and $A_{\Gamma}\neq 0$ indicate indirect CPV. $A_{\Gamma}$ is known to be very small. We also define~\cite{PDG}
\begin{equation}
\begin{split}
&\left|\frac{q}{p}\right|^2\equiv \sqrt{\frac{1+A_M}{1-A_M}},
\end{split}
\label{eq.4.11}
\end{equation}
which is often used in the studies of $D$ decays.

\subsection{\label{sec4.1} TV signals based on joint decay rates}

For the $C=-1$ entangled state, we can construct four TV signals from time-dependent joint decay rates (depending on the difference $\Delta t =t_b-t_a$ of two decay times), corresponding to the final states listed in Table~{\ref{tab1}}. In the first example listed in the Table~{\ref{tab1}}, the final states of mesons $a$ and $b$ are  $l^-$ and $S_-$, with direct CPV neglected, i.e. $A_{l^-}=\bar{A}_{l^+}=0$, $A_{l^+}=\bar{A}_{l^-}\equiv A_l$, $A_{S_{\pm}}=\pm \bar{A}_{S_{\pm}}$,  $ R_-(l^-,S_-,t_a,t_b)\propto |A_l|^2|A_{S_-}|^2|\langle D_-|U(t_b-t_a)|D^0\rangle|^2$. In details, if the final state of meson $a$ is $f_a$,  the state of meson $b$ becomes
$ \propto A_{f_a} |\bar{D}^0\rangle- \bar{A}_{f_a}|D^0\rangle$.
Then, if the meson $b$ decays into $f_b$ at $t_b$,  it can be obtained that
\begin{equation}
\begin{split}
&|\langle f _a, f _b|{\cal H}_a{\cal H}_b|\Psi _-(t_a,t_b)\rangle |^2 \\ &
 = e^{-2\Gamma t_a} \left( |A_{f_a}|^2|\langle f_b|{\cal H}_bU(t_b-t_a)|\bar{D}^0\rangle|^2+|\bar{A}_{f_a}|^2|\langle f_b|{\cal H}_bU(t_b-t_a)|D^0\rangle|^2 \right)\\
& = e^{-2\Gamma t_a}  ( \frac{1}{2}|A_{f_a}|^2(|\langle f_b|{\cal H}_bD_+\rangle|^2|\langle D_+|U(t_b-t_a)|\bar{D}^0\rangle|^2+|\langle f_b|{\cal H}_bD_-\rangle|^2|\langle D_-|U(t_b-t_a)|\bar{D}^0\rangle|^2)  \\
&  +\frac{1}{2}|\bar{A}_{f_a}|^2(|\langle f_b|{\cal H}_b|D_+\rangle|^2|\langle D_+|U(t_b-t_a)|D^0\rangle|^2+|\langle f_b|{\cal H}_b|D_-\rangle|^2|\langle D_-|U(t_b-t_a)|D^0\rangle|^2) ),
\end{split}
\end{equation}
where we have assumed no wrong-sign decay. If $f_a=l^-$, then
\begin{equation}
\begin{split}
& |\langle f _a, f _b|{\cal H}_a{\cal H}_b|\Psi_-(t_a,t_b)\rangle |^2 \\
& = \frac{ e^{-2\Gamma t_a} }{2} \left( |A_l|^2 (|\langle f_b|{\cal H}_b|D_+\rangle|^2|\langle D_+|D^0(t_b-t_a)\rangle|^2+|\langle f_b|{\cal H}_b|D_-\rangle|^2 |\langle D_-|D^0(t_b-t_a)\rangle|^2)\right).
\end{split}
\end{equation}
Considering
$\langle S_-|{\cal H}|D_+\rangle=\frac{\sqrt{2}}{2} (\langle S_-|{\cal H}|D^0\rangle+\langle S_-|{\cal H}|\bar{D}^0\rangle )=0$, $\langle S_-|{\cal H}|D_-\rangle=\frac{\sqrt{2}}{2}A_{S_-}$, and $f_b=S_-$,  we have
\begin{equation}
|\langle l^-, S_-|{\cal H}_a{\cal H}_b|\Psi _-(t_a,t_b)\rangle |^2=\frac{ e^{-2\Gamma t_a}}{4}  |A_l|^2|A_{S_-}|^2|\langle D_-|D^0(t_b-t_a)\rangle|^2.
\label{l-1}
\end{equation}
As $|D^0(t)\rangle=g_+(t)|D^0\rangle-\frac{q}{p}g_-(t)|\bar{D}^0\rangle$, it can be obtained that
\begin{equation}
\begin{split} &|\langle D_-|D^0(t_b-t_a)\rangle|^2 =\frac{1}{4}e^{-\Gamma (t_a-t_b)} e^{-\frac{\Delta \Gamma}{2} (t_a-t_b)}  \left((\frac{q}{p}+1) e^{\frac{1}{2} (t_a-t_b)(\Delta \Gamma+2 i \Delta m)}-\frac{q}{p}+1\right)\\& \times \left(\frac{q}{p} \left(-1+e^{\left(\frac{1}{4}+\frac{i}{4}\right) (t_a-t_b) (\Delta \Gamma-2 i (\Delta m+\Gamma)-4 m)}\right)+e^{\frac{1}{2} (t_a-t_b) (\Delta \Gamma-2 i \Delta m)}+1\right).\end{split}\end{equation}
Hence, Eq.~(\ref{l-1}) is consistent with Eq.~(\ref{eq.3.9}); especially, $|\langle l^-, S_-|{\cal H}_a{\cal H}_b|\Psi_-(t_a,t_b)\rangle |^2 \propto e^{ - \Gamma (t_a+t_b) }$.

Similarly,
\begin{equation}
R_-(S_+,l^+,t_a,t_b)=\frac{ e^{-2\Gamma t_a}}{4}
|A_{S_+}|^2 |A_l|^2 \langle D^0|U(t_b-t_a)|D_-\rangle|^2,
\end{equation}
A similar expression can be for each pair of T-conjugated transitions.

T symmetry implies $|\langle D_-|U(\Delta t)|D^0\rangle |^2=|\langle D^0|U(\Delta t)|D_-\rangle |^2$. Therefore,  for $\Delta t>0$, T symmetry implies that
\begin{equation}
\begin{split}
&\frac{R_-(l^-,S_-,\Delta t)}{|A_l|^2|A_{S_-}|^2}=\frac{R_-(S_+,l^+,\Delta t)}{|A_l|^2|A_{S_+}|^2},\;\;\frac{R_-(l^-,S_+,\Delta t)}{|A_l|^2|A_{S_+}|^2}=\frac{R_-(S_-,l^+,\Delta t)}{|A_l|^2|A_{S_-}|^2},\\
&\frac{R_-(l^+,S_-,\Delta t)}{|A_l|^2|A_{S_-}|^2}=\frac{R_-(S_+,l^-,\Delta t)}{|A_l|^2|A_{S_+}|^2},\;\;\frac{R_-(l^+,S_+,\Delta t)}{|A_l|^2|A_{S_+}|^2}=\frac{R_-(S_-,l^-,\Delta t)}{|A_l|^2|A_{S_-}|^2}.\\
\end{split}
\label{eq.4.8.add1}
\end{equation}
Hence, we can define a T asymmetry, denoted as $A_-^1(\Delta t > 0)$,
\begin{equation}
\begin{split}
&A_-^1(\Delta t ) = \frac{\frac{R_-(l^-,S_-,\Delta t)}{|A_l|^2|A_{S_-}|^2}-\frac{R_-(S_+,l^+,\Delta t)}{|A_l|^2|A_{S_+}|^2}}{\frac{R_-(l^-,S_-,\Delta t)}{|A_l|^2|A_{S_-}|^2}+\frac{R_-(S_+,l^+,\Delta t)}{|A_l|^2|A_{S_+}|^2}}, \\
\end{split}
\label{eq.4.8.add2}
\end{equation}
and there are three other asymmetries corresponding to the equalities in Eq.~(\ref{eq.4.8.add1}).

We can also define TV signals   independent of $|A_{S_{\pm}}|$,   denoted as $A_-^2(\Delta t > 0)$,
\begin{equation}
\begin{split}
&A_-^2(\Delta t ) = \frac{R_-(l^-,S_-,\Delta t)}{R_-(l^+,S_-,\Delta t)}-\frac{R_-(S_+,l^+,\Delta t)}{R_-(S_+,l^-,\Delta t)},\\
\end{split}
\label{eq.4.8.add3}
\end{equation}
There are five other signals similar to Eq.~(\ref{eq.4.8.add3}) that can be constructed, according to Eq.~(\ref{eq.4.8.add1}).

One can also use the normalized joint decay rates  or the probability density function~(PDF),  defined as
\begin{equation}
r_-(f_a,f_b,\Delta t)=\frac{1}{n_{f_a,f_b}}\frac{R_-(f_a,f_b,\Delta t)}{|A_{f_a}|^2|A_{f_b}|^2}= \frac{1}{n'_{f_a,f_b}} R_-(f_a,f_b,\Delta t),
\label{eq.4.8.add4}
\end{equation}
where $n_{f_a,f_b}=\int _0^{\infty}d(\Delta t)\frac{R_-(f_a,f_b,\Delta t)}{|A_{f_a}|^2|A_{f_b}|^2}$, $n'_{f_a,f_b}=\int _0^{\infty}d(\Delta t)R_-(f_a,f_b,\Delta t)$.  That is to say, the PDF  for $R_-(f_a,f_b,\Delta t)$ is the same as that   for $\frac{R_-(f_a,f_b,\Delta t)}{|A_{f_a}|^2|A_{f_b}|^2}$. Therefore, one only needs to consider  $R_-(f_a,f_b,\Delta t)$ when normalization with respect to various  $\Delta t$ is taken into account. Hence, one can construct a TV  $A_-^3(\Delta t > 0)$ as
\begin{equation}
\begin{split}
&A_-^3(\Delta t ) = \frac{r_-(l^-,S_-,\Delta t)-r_-(S_+,l^+,\Delta t)}{r_-(l^-,S_-,\Delta t)+r_-(S_+,l^+,\Delta t)},\\
\end{split}
\label{eq.4.8}
\end{equation}
which vanishes only if T symmetry is valid. Note that it was  $A_-^3(\Delta t)$ that  was measured in Barbar experiments~\cite{TVB,mcmethod}.

We now consider the  time-independent joint decay rate
\begin{equation}
\begin{split}
&R_-(f_a,f_b) =\int _0^{\infty}dt_a\int _0^{\infty}dt_b R_-(f_a,f_b,t_a,t_b)=\int _0^{\infty}dt_a\int _0^{\infty}dt_b R_-(f_a,f_b,t_b,t_a).\\
\end{split}
\label{eq.4.16}
\end{equation}
Note that $R_-(f_a,f_b)/|A_{f_a}|^2|A_{f_b}|^2=R_-(f_b,f_a)/|A_{f_a}|^2|A_{f_b}|^2$ is independent of the order of the final states. Hence in counting the events, one does not need to distinguish which final state is of which meson.

$R_-(l^-,S_-)/|A_l|^2|A_{S_-}|^2 \neq R_-(l^+,S_+)/|A_l|^2|A_{S_+}|^2$ is a sufficient condition of TV in the time-dependent rates and implies that there is at least a certain value of $t$, for which at least one of the two corresponding conjugate processes    violates T symmetry. A similar conclusion can be made if $R_-(l^-,S_+)/|A_l|^2|A_{S_+}|^2 \neq R_-(l^+,S_-)/|A_l|^2|A_{S_-}|^2$.

If time reversal symmetry is respected, then  both of the following equations are satisfied
\begin{equation}
\begin{split}
&\frac{R_-(l^-,S_-)}{|A_l|^2|A_{S_-}|^2} = \frac{R_-(l^+,S_+)}{|A_l|^2|A_{S_+}|^2},\;\;\;\;\frac{R_-(l^-,S_+)}{|A_l|^2|A_{S_+}|^2} = \frac{R_-(l^+,S_-)}{|A_l|^2|A_{S_-}|^2}.
\end{split}
\label{eq.4.17.add}
\end{equation}
Hence we can define the time-independent TV signal of $C=-1$ states denoted as $\hat{A}_-$,
\begin{equation}
\begin{split}
&\hat{A}_-=\frac{R_-(l^-,S_-)}{R_-(l^+,S_-)}-\frac{R_-(l^+,S_+)}{R_-(l^-,S_+)}.
\end{split}
\label{eq.4.17}
\end{equation}
When $\hat{A}_-\neq 0$, at least one of the equalities  in Eq.~(\ref{eq.4.17.add}) is violated. Therefore  $\hat{A}_-$ is the TV signal independent of $A_{S_{\pm}}$.

We emphasize that $A_-^2(\Delta t)= 0$ or $A_-^3(\Delta t)= 0$ or $A_-= 0$ does not guarantee the time reversal symmetry. However,  $A_-^2(\Delta t)\neq 0$ or $A_-^3(\Delta t)\neq 0$ or  $A_-\neq 0$ is a sufficient condition of TV. In experiments, one would like to use the TV signal independent of $A_{S_{\pm}}$, that is, $A_-^2(\Delta t)$, $A_-^3(\Delta t)$ and $\hat{A}_-$.

Note that, despite the decays, the antisymmetry of the $C=-1$ entangled state remains. This is crucial in its use in the construction of genuine TV signals~\cite{Treview1}. The $C=+1$ entangled state of $D$ mesons can also be produced in the strong decay of $\psi(4140)$~\cite{JDRxzz1,JDRxzz2}, but it is difficult to extract TV signals from it.  When he $C=+1$ entangled state evolves to $t=t_a$, it becomes $|\Psi _+(t_a)\rangle$ as given in (\ref{plus}). Consequently,  when one of the mesons decays into the $f_a$ final state at $t_a$, the other meson becomes a superposition of $D^0$ and $\bar{D}^0$. If we denote $\Psi _{f_a}$ as the state of the second meson tagged by the final state of the first meson $f_a$, $\Psi _{f_a}$ can be written as
\begin{equation}
\begin{split}
&|\Psi _{l^+}\rangle \propto \frac{q}{p} \cos(\Delta \lambda  t_a) |\bar{D}^0\rangle + i\frac{1+\left(\frac{q}{p}\right)^2}{2}\sin(\Delta \lambda t_a) |D^0\rangle  +i\frac{1-\left(\frac{q}{p}\right)^2}{2}\sin(\Delta \lambda t_a) |D^0\rangle,\\
&|\Psi _{l^-}\rangle \propto \frac{q}{p} \cos(\Delta \lambda  t_a) |D^0\rangle + i\frac{1+\left(\frac{q}{p}\right)^2}{2}\sin(\Delta \lambda t_a) |\bar{D}^0\rangle  -i\frac{1-\left(\frac{q}{p}\right)^2}{2}\sin(\Delta \lambda t_a) |\bar{D}^0\rangle,\\
&|\Psi _{S_+}\rangle \propto \frac{q}{p} \cos(\Delta \lambda  t_a)|D_+\rangle +i\frac{1+\left(\frac{q}{p}\right)^2}{2}\sin(\Delta \lambda t_a) |D_+\rangle +i\frac{1-\left(\frac{q}{p}\right)^2}{2}\sin(\Delta \lambda t_a) |D_-\rangle,\\
&|\Psi _{S_-}\rangle \propto -\frac{q}{p} \cos(\Delta \lambda  t_a)|D_-\rangle +i\frac{1+\left(\frac{q}{p}\right)^2}{2}\sin(\Delta \lambda t_a) |D_-\rangle
+ i\frac{1-\left(\frac{q}{p}\right)^2}{2}\sin(\Delta \lambda t_a) |D_+\rangle,\\
\end{split}
\end{equation}
where $\propto$ implies that these four states are not normalized yet.
If, for example,  we compare the joint decay rate $R_+(l^-, S_-, t_a, t_b)$ with $R_+(S_+, l^+, t_a, t_b)$, we are comparing the transitions $\Psi _{l^-}\to D_-$ with $\Psi _{S_+}\to D^0$, which are not T-conjugate transitions.

In the following, we concentrate on the TV signals of the $C=-1$ entangled states. Substituting Eq.~(\ref{eq.4.4}) into Eq.~(\ref{eq.3.13}), we obtain  the  time-dependent joint decay rates
\begin{equation}
\begin{split}
&R_-(l^+, S_{\pm}, \Delta t)= \frac{e^{-\Gamma |\Delta t|} |A_l|^2 |A_{S_{\pm}}|^2}{8\Gamma} \left\{\left(\left|\frac{p}{q}\right|^2+1\right)\cosh (y\Gamma \Delta t)-\left(\left|\frac{p}{q}\right|^2-1\right)\cos (x\Gamma \Delta t)\right.\\
&\left. \mp 2\left|\frac{p}{q}\right| \left[\cos (2\phi)\sinh (y\Gamma \Delta t)+\sin (2\phi)\sin (x \Gamma \Delta t)\right]\right\},\\
&R_-(l^-, S_{\pm}, \Delta t)= \frac{e^{-\Gamma |\Delta t|} |A_l|^2 |A_{S_{\pm}}|^2}{8\Gamma} \left\{\left(\left|\frac{q}{p}\right|^2+1\right)\cosh (y\Gamma \Delta t)-\left(\left|\frac{q}{p}\right|^2-1\right)\cos (x\Gamma \Delta t)\right.\\
&\left. \mp 2\left|\frac{q}{p}\right|\left[\cos (2\phi)\sinh (y\Gamma \Delta t)-\sin (2\phi)\sin (x \Gamma \Delta t)\right]\right\}.\\
\end{split}
\label{eq.4.5}
\end{equation}

With $\Lambda \equiv -q/p \times \bar{A}_{S_{\pm}} / A_{S_{\pm}}$, and at the limit at which $\Delta \Gamma \to 0$, which is the case of $B$ mesons~\cite{mcmethod}, the integrated joint decay rates become
\begin{equation}
\begin{split}
&R_{-}(l^{+}, S_{\pm}, \Delta t)|_B \propto e^{-\Gamma |\Delta t|} \left(1 - \left(\frac{1-|\Lambda|^2}{1+|\Lambda|^2}\cos (x\Gamma \Delta t) - \frac{2{\rm Im}\Lambda}{1+|\Lambda| ^2}\sin (x \Gamma \Delta t)\right)\right),\\
&R_{-}(l^{-}, S_{\pm}, \Delta t)|_B \propto e^{-\Gamma |\Delta t|} \left(1+ \left(\frac{1-|\Lambda|^2}{1+|\Lambda|^2}\cos (x\Gamma \Delta t) - \frac{2{\rm Im}\Lambda}{1+|\Lambda| ^2}\sin (x \Gamma \Delta t)\right)\right),
\end{split}
\label{eq.4.6}
\end{equation}
which reproduces the integrated joint decay rates of $B$ mesons in Refs.~\cite{Babar2007,Babar2009}.

The time-independent joint decay rate can be obtained, from Eqs.~(\ref{eq.3.15}) and (\ref{eq.4.4}):
\begin{equation}
\begin{split}
&R_-(l^+, S_{\pm})= \frac{|A_l|^2 |A_{S_{\pm}}|^2}{4\Gamma^2}\left\{\left(\left|\frac{p}{q}\right|^2+1\right)\frac{1}{1-y^2}-\left(\left|\frac{p}{q}\right|^2-1\right)\frac{1}{1+x^2}\right\},\\
&R_-(l^-, S_{\pm})= \frac{|A_l|^2 |A_{S_{\pm}}|^2}{4\Gamma^2}\left\{\left(\left|\frac{q}{p}\right|^2+1\right)\frac{1}{1-y^2}-\left(\left|\frac{q}{p}\right|^2-1\right)\frac{1}{1+x^2}\right\}.\\
\end{split}
\label{eq.4.7}
\end{equation}

Now we can obtain the TV signals.
Taking $A_-^1(\Delta t)$ as an example, we can estimate $A_-^1(\Delta t)$ of the $C=-1$ system using the measured parameters of CPV of $D^0-\bar{D}^0$ mesons in the SM. Using Eqs.~(\ref{eq.4.8}) and (\ref{eq.4.5}), we find
\begin{equation}
\begin{split}
&A_-^1(\Delta t) =\frac{ \left(x_1y_1(\Delta t)-2x_2\cos (2\phi)\sinh (y\Gamma \Delta t)-2x_3 \sin (2\phi)\sin (x\Gamma \Delta t)\right) }
{ \left(x_4y_1(\Delta t)+2y_2(\Delta t)+2x_3\cos (2\phi)\sinh (y\Gamma \Delta t)+2x_2\sin(2\phi)\sin(x\Gamma \Delta t)\right) },
\end{split}
\label{eq.4.9}
\end{equation}
where $x_i$ and $y_i$ are defined as
\begin{equation}
\begin{split}
&x_1\equiv\left|\frac{q}{p}\right|^2-\left|\frac{p}{q}\right|^2,\;
 x_2\equiv\left|\frac{p}{q}\right|-\left|\frac{q}{p}\right|,\;
 x_3\equiv\left|\frac{p}{q}\right|+\left|\frac{q}{p}\right|,\;
 x_4\equiv\left|\frac{p}{q}\right|^2+\left|\frac{q}{p}\right|^2,\\
&y_1(\Delta t)\equiv\cosh (y\Gamma \Delta t)-\cos (x \Gamma \Delta t),\;
 y_2(\Delta t)\equiv\cosh (y\Gamma \Delta t)+\cos (x \Gamma \Delta t).
\end{split}
\label{eq.4.xdefine}
\end{equation}

In the case of $B$ mesons, we can take the limit $\Delta \Gamma \to 0$ and $q / p \to e^{2i\beta}$; thus, we find $A_-^1(\Delta t)=-\sin (2\beta)\sin (x\Gamma \Delta t)$. This corresponds to the $CP$ asymmetry predicted by the SM, as given in Refs.~\cite{Babar2007,Babar2009}.

We can expand $A_-^i(\Delta t)$ to the leading order and find
\begin{equation}
\begin{split}
&A_-^1(\Delta t)\approx A_{\Gamma}\Gamma \Delta t,\;\;A_-^2(\Delta t)\approx 4A_{\Gamma}\Gamma \Delta t,\;\;A_-^3(\Delta t)\approx A_{\Gamma}\Gamma (\Delta t-1).\\
\end{split}
\label{eq.4.12}
\end{equation}
We use the parameter values in Ref.~\cite{HFAG},
\begin{equation}
\begin{split}
&x=0.0037,\;\;y=0.0066,\;\;\frac{q}{p}=0.91,
\;\;\phi=-4.7^o .
\end{split}
\label{eq.4.13}
\end{equation}
Notice that the definition of $\phi$ in Ref.~\cite{HFAG} is $\arg\left(q/p\right)$, while in this paper, we define $\phi \equiv \arg\left(q/p\right)/2$, which is the same as in Ref.~\cite{JDRxzz1}. For $\Delta t = \tau_D \equiv 1/\Gamma $, we find $A_-^1(\Delta t) \sim 10^{-5}$.

The time-independent joint decay rate does not depend on the decay times, so we are not able to identify the transition. For example, we need to know which of the final states is the outcome of the earlier decay to distinguish $D^0\to D_-$ from $D_+\to \bar{D}^0$. However, one can construct a time-independent signal for TV.

It is found that
\begin{equation}
\begin{split}
&\hat{A}_-=\frac{x_1  (x_4(\bar{x}-\bar{y})^2+2 (\bar{x}^2-\bar{y}^2))}{x_4 \left(\bar{x}^2-\bar{y}^2\right)+2 \left(\bar{x}^2+\bar{y}^2\right)},\\
&\bar{x}\equiv 1+x^2,\;\;\bar{y}\equiv 1-y^2,\\
\end{split}
\label{eq.4.18}
\end{equation}
where $x_1$ and $x_4$ are defined in Eq.~(\ref{eq.4.xdefine}). To the leading order,
\begin{equation}
\begin{split}
&\hat{A}_- \approx 2A_M(x^2+y^2)= -2.2\times 10^{-5}.
\end{split}
\label{eq.4.19}
\end{equation}
The error  of the signal can be estimated to be related to the event number $N$ as $\delta \hat{A}_- \sim 1/\sqrt{N}$. Hence the magnitude of $\hat{A}_-$ implies that the number of events should be as large as $10^9$ to $    10^{10}$,  which will be verified in Monte Carlo simulation in Sec.~\ref{sec:level6}. Such an event number can be obtained at the super-tau-charm factory~\cite{supertaucharm}.

\subsection{\label{sec4.3}\texorpdfstring{$C=-1$}{C=-1} state with \texorpdfstring{$\omega$ effect}{omega effect}}

As noted in Eq.~(\ref{eq.3.16}), the $\omega$ effect causes the $C=-1$ state to be mixed in by  the $C=+1$ state. Then the T-conjugation between each pair of processes  in the asymmetries studied above is lost. However, the asymmetries for these pairs of processes can still be investigated to determine the value of $\omega$.
We find that these asymmetries are  enhanced. For example, for the same final states as in $A_-^1(\Delta t)$ defined in Eq.~(\ref{eq.4.8.add2}), the corresponding asymmetry of the $C=+1$ state is
\begin{equation}
\begin{split}
&A_+^1(\Delta t ) =\frac{\frac{R_+(l^-,S_-,\Delta t)}{|A_l|^2|A_{S_-}|^2}-\frac{R_+(S_+,l^+,\Delta t)}{|A_l|^2|A_{S_+}|^2}}{\frac{R_+(l^-,S_-,\Delta t)}{|A_l|^2|A_{S_-}|^2}+\frac{R_+(S_+,l^+,\Delta t)}{|A_l|^2|A_{S_+}|^2}}.\\
\end{split}
\label{eq.4.enhanced}
\end{equation}

Inserting Eq.~(\ref{eq.4.4}) into Eqs.~(\ref{eq.3.13}) and (\ref{eq.3.15}), in the case of $C=+1$, we find
\begin{equation}
\begin{split}
&A_+^1(\Delta t)\approx y_{\rm CP}(1+\Gamma \Delta t)\sim 10^{-2},\\
\end{split}
\label{eq.4.enhancedsignal}
\end{equation}
The difference between $A_-^1(\Delta t)$ and $A_+^1(\Delta t)$ is very large, providing an opportunity to detect the $\omega$ effect. The numerical results show that $A_-^1(\Delta t\approx \tau _D)/A_+^1(\Delta t \approx \tau _D) \sim 10^{-4}$, which implies that a small $\omega$ at the order $|\omega|\sim 10^{-4}$ may considerably change the TV signals. Incidently, this is also the order of magnitude considered in Ref.~\cite{omega}.    So we conjecture the experiment to observe the TV signal in the $D$ system may at the same time provide a window to detect the $\omega$ effect with a sensibility up to $|\omega|\sim 10^{-4}$.

For simplicity, we only consider how the TV signal $A_-^2(\Delta t)$ is affected by the $\omega $ effect. Using Eqs.~(\ref{eq.3.7}), (\ref{eq.3.19})-(\ref{eq.4.2}),  we find
\begin{equation}
\begin{split}
&A_{\omega}(\Delta t)= \frac{R_{\omega}(l^-,S_-,\Delta t)}{R_{\omega}(l^+,S_-,\Delta t)}-\frac{R_{\omega}(S_+,l^+,\Delta t)}{R_{\omega}(S_+,l^-,\Delta t)}\\
&\approx \frac{1}{4} \Gamma \Delta t \left(-2 \sin(2\phi) x \left(A_M^2-8y \cos(2\phi) \Gamma \Delta t+8\right)+A_M \left(3 A_M^2+8\right) \cos(2\phi) y\right.\\
&\left.+4 A_M\Gamma \Delta t \left(\left(1-2 \cos^2(2\phi)\right) y^2+x^2\right)\right)+4 \cos(2\phi)|\omega|(y \cos (\Omega)-x \sin (\Omega))(1+\Gamma \Delta t),
\end{split}
\label{eq.4.26}
\end{equation}
where $A_M$ is determined by $q/p$, as defined in Eq.~(\ref{eq.4.11}).

\begin{figure}
\resizebox{0.55\textwidth}{!}{%
\includegraphics{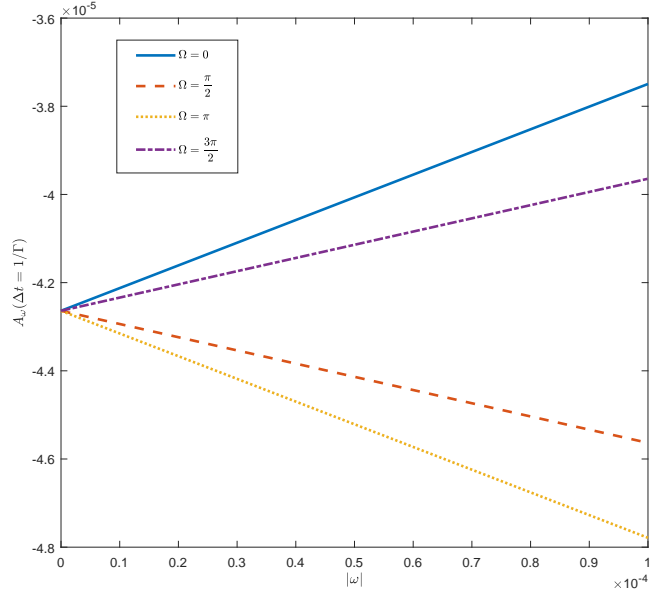}
}
\caption{\label{Fig:amtomi} $A_{\omega}(\Delta t=1/\Gamma)$ as a function  of $|\omega|$ in the region $|\omega|<10^{-4}$. The solid line is for $\Omega=0$, the dashed line is for $\Omega=\pi / 2$, the dotted line is for $\Omega=\pi$, and the dotted-dashed line is for $\Omega = 3\pi/2$. The parameter values are $x=0.0037$, $y=0.0066$, $\frac{q}{p}=0.91$, and $\phi=-4.7^o$. }
\end{figure}

\begin{figure}
\resizebox{0.55\textwidth}{!}{%
\includegraphics{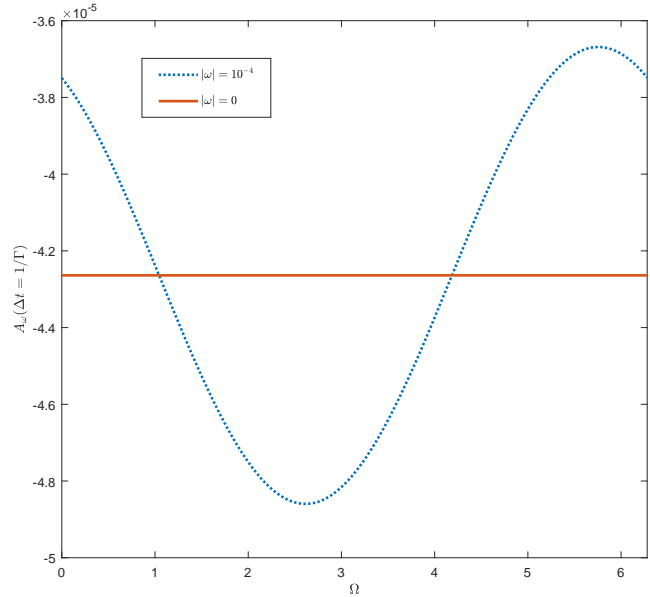}
}
\caption{\label{Fig:amtOm} $A_{\omega}(\Delta t=1/\Gamma)$ as a function  of $\Omega$. The solid line is for $|\omega|=0$, that is, within the SM. The dotted line is for $|\omega|=10^{-4}$.  The parameter values are $x=0.0037$, $y=0.0066$, $\frac{q}{p}=0.91$, and $\phi=-4.7^o$.}
\end{figure}

The CPV parameters are assumed to be barely affected by the $\omega$ effect. Using Eq.~(\ref{eq.4.13}), the dependence of $A_{\omega}(\Delta t)$ on $|\omega|$ and $\Omega$ when $\Gamma \Delta t =1$, i.e., $\Delta t = \tau_D \equiv 1/\Gamma$,  is shown in Figs.~\ref{Fig:amtomi} and \ref{Fig:amtOm}. We find that when $|\omega|\sim 10^{-4}$ the change of time-integrated T asymmetry, due to the $\omega$ effect, can be as large as $20\%$ of that within the SM.  The sensitivity could be competitive with the $B$ or $B_d$ meson pairs~\cite{omegabd}. In  the Monte Carlo simulation presented in  Sec.~\ref{sec:level6}, we  will find that if the event number is of the order of  $10^9$ the TV signal can possibly be observed. Such an event number can also set a bound on $|\omega|$ at $10^{-3}$ at the same time.

We emphasize when the $C=-1$ state is mixed with the $C=+1$ state the signal is no longer a TV signal. However, the deviation from the TV signal calculated within the SM  reveals the nonzero $\omega$ effect.

\section{\label{sec:level5}Relation between the TV signals and experimental measurements}

One can relate the normalized time-integrated joint decay rates to event numbers of the decays~\cite{mcmethod}. In using normalized time-integrated joint decay rates,  the T-conjugated transitions differ in the dependence on the time interval rather than on the number of events.

A  similar way to investigate the double decay is to use ST  and  DT  signals~\cite{DT,DT2,BESCII}.

Suppose the final state of meson $a$ at $t_a$ is $l^-$, it tagged the meson $b$ as $D^0$, which decays to $S_-$ at $t_b=t_a+\Delta t$, the rate of which can be denoted as $\Gamma (D^0\to S_-,\Delta t)$. By assuming that there is no mistake in tagging and that the direct CPV can be neglected, the rate $|\langle D_-|U(\Delta t)|D^0\rangle |^2$ of the transition $D^0\to D_-$ in time interval $\Delta t$ is related to decay rate $\Gamma (D^0\to S_-,\Delta t)$ as
\begin{equation}
\begin{split}
&\Gamma (D^0\to S_-,\Delta t)\propto |\langle S_-|\mathcal{H}|D^0(\Delta t)\rangle|^2\\
&=|\langle S_- |\mathcal{H}| D_-\rangle \langle D_-|U(\Delta t)|D^0\rangle +\langle S_- |\mathcal{H}| D_+\rangle \langle D_+|U(\Delta t)|D^0\rangle|^2\\
&=|\langle D_-|U(\Delta t)|D^0\rangle |^2 | \langle S_- |\mathcal{H}| D_-\rangle|^2,
\end{split}
\end{equation}
where $\mathcal{H}$ is the Hamiltonian governing the decay. As a result
\begin{equation}
\begin{split}
&\Gamma (D^0\to S_-,\Delta t)= |\langle D_-|U(\Delta t)|D^0\rangle |^2|\Gamma (D_-\to S_-),
\end{split}
\end{equation}
where $\Gamma (D_-\to S_-) \equiv  | \langle S_- |\mathcal{H}| D_-\rangle|^2$.

In experiments, the decay rate can be related to event numbers as
\begin{equation}
\begin{split}
&N_{l^-,S_-}(t_a,t_a+\Delta t_0)=\int _0^{\Delta t_0}\Gamma (D^0\to S_-,\Delta t)N_{l^-}(t_a)d(\Delta t)
\end{split}
\end{equation}
where $N_{l^-}(t_a)$ is the number of the events in which  meson $a$ decays to $l^-$ at $t_a$ and $N_{l^-,S_-}(t_a,t_a+\Delta t_0)$ is the number of the joint events in which meson $a$ decays to $l^-$ at $t_a$ and then meson $b$ decays to $S_-$ in time interval $[t_a, t_a+\Delta t_0]$. So
\begin{equation}
\begin{split}
&\int _0^{\infty}dt_a N_{l^-,S_-}(t_a,t_a+\infty)=\int _0^{\infty}\Gamma (D^0\to S_-,\Delta t)d(\Delta t)\int _0^{\infty}N_{l^-}(t_a)dt_a,
\end{split}
\label{eq.3.nint}
\end{equation}
which can be rewritten as
\begin{equation}
\begin{split}
&{\cal N}_{l^-,S_-}=R (D^0\to S_-){\cal N}_{l^-},\\
\end{split}
\label{eq.3.singlerateint}
\end{equation}
where
\begin{equation}
\begin{split}
&{\cal N}_{l^-,S_-}\equiv \int _0^{\infty}dt_a  N_{l^-,S_-}(t_a,t_a+\infty),\;\;
 {\cal N}_{l^-}\equiv \int _0^{\infty}N_{l^-}(t_a)dt_a,\\
&R (D^0\to S_-)\equiv \int _0^{\infty}\Gamma (D^0\to S_-,\Delta t)d(\Delta t)
=R (D^0\to D_-)\Gamma (D_-\to S_-),
\end{split}
\end{equation}
with
\begin{equation}
R (D^0\to D_-)\equiv \int _0^{\infty}|\langle D_-|U(\Delta t)|D^0\rangle |^2d(\Delta t).
\end{equation}
${\cal N}_{l^-}$ is the total number of events in which  meson $a$ decays to $l^-$  and  is also called the signal yield of ST decays. ${\cal N}_{l^-,S_-}$ is the the total number of  the  joint events in which meson $a$ decays to $l^-$ while meson $b$ decays to $S_-$ and is also called the signal yield of DT decays.

Since T symmetry requires $|\langle D_-|U(\Delta t)|D^0\rangle |^2=|\langle D^0|U(\Delta t)|D_-\rangle |^2$ for any $\Delta t>0$, $R(D^0\to D_-)\neq R(D_-\to D^0)$ is a sufficient TV signal.

In experiments, the detection efficiencies should also be considered, so we can write  the transition rates as
\begin{equation}
\begin{split}
&R(D^0\to S_{\pm})\equiv R(D^0\to D_{\pm})\Gamma(D_{\pm}\to S_{\pm}) = \frac{{\cal N}_{l^-,S_{\pm}}}{{\cal N}_{l^-}}\frac{\varepsilon _{l^-}}{\varepsilon _{l^-,S_{\pm}}},\\
&R(\bar{D}^0\to S_{\pm})\equiv R(\bar{D}^0\to D_{\pm})\Gamma(D_{\pm}\to S_{\pm}) = \frac{{\cal N}_{l^+,S_{\pm}}}{{\cal N}_{l^+}}\frac{\varepsilon _{l^+}}{\varepsilon _{l^+,S_{\pm}}},\\
&R(D_{\pm}\to l^+)\equiv R(D_{\pm}\to D^0)\Gamma(D^0\to l^+) =\frac{{\cal N}_{S_{\mp},l^+}}{{\cal N}_{S_{\mp}}}\frac{\varepsilon _{S_{\mp}}}{\varepsilon _{S_{\mp},l^+}},\\
&R(D_{\pm}\to l^-)\equiv R(D_{\pm}\to \bar{D}^0)\Gamma(\bar{D}^0\to l^-) = \frac{{\cal N}_{S_{\mp},l^-}}{{\cal N}_{S_{\mp}}}\frac{\varepsilon _{S_{\mp}}}{\varepsilon _{S_{\mp},l^-}},\\
\end{split}
\label{eq.3.signalsbyexperiments}
\end{equation}
where $\varepsilon$'s are the detection efficiencies, with the subscripts the same as those of the corresponding event numbers ${\cal N}$'s, which are now understood as the experimental  ones.

If time reversal symmetry is conserved, $R(D^0\to D_-)=R(D_-\to D^0)$, $R(\bar{D}^0\to D_-)=R(D_-\to \bar{D}^0)$. Then according to Eq.~(\ref{eq.3.signalsbyexperiments}), we have
\begin{equation}
\begin{split}
&\frac{R(D^0\to S_-)}{\Gamma(D_-\to S_-)}=\frac{R(D_-\to l^+)}{\Gamma(D^0\to l^+)},\;
 \frac{R(\bar{D}^0\to S_-)}{\Gamma(D_-\to S_-)}=\frac{R(D_-\to l^-)}{\Gamma(\bar{D}^0\to l^-)}.\\
\end{split}
\label{eq.4.singletransitiontimeconserve}
\end{equation}

By using the ratios between the left-hand sides and right-hand sides of the equalities in Eq.~(\ref{eq.4.singletransitiontimeconserve}),   we construct the TV signal $A_T^1$ as
\begin{equation}
\begin{split}
&A_T^1=\frac{\Gamma(D_-\to S_-)}{\Gamma(D_-\to S_-)}\frac{R(D^0\to S_-)}{R(\bar{D}^0\to S_-)}
 -\frac{\Gamma(\bar{D}^0\to l^-)}{\Gamma(D^0\to l^+)}\frac{R(D_-\to l^+)}{R(D_-\to l^-)}\\
&=\frac{\frac{{\cal N}_{l^-,S_-}}{{\cal N}_{l^-}}\frac{\varepsilon _{l^-}}{\varepsilon _{l^-,S_-}}}{\frac{{\cal N}_{l^+,S_-}}{{\cal N}_{l^+}}\frac{\varepsilon _{l^+}}{\varepsilon _{l^+,S_-}}}
 -\frac{\Gamma(\bar{D}^0\to l^-)}{\Gamma(D^0\to l^+)}\frac{\frac{{\cal N}_{S_+,l^+}}{{\cal N}_{S_+}}\frac{\varepsilon _{S_+}}{\varepsilon _{S_+,l^+}}}{\frac{{\cal N}_{S_+,l^-}}{{\cal N}_{S_+}}\frac{\varepsilon _{S_+}}{\varepsilon _{S_+,l^-}}},
\end{split}
\label{eq.4.dtsignal}
\end{equation}
which can thus be obtained from the numbers of ST and DT events. Here, $A_T^1\neq 0$ is a TV signal. Note that  $A_T^1=0$ does not guarantee   T symmetry; however, $A_T^1\neq 0$ is a sufficient condition of TV.

Another T-asymmetry can be constructed as
\begin{equation}
\begin{split}
&A_T^2=\frac{R(\bar{D}^0\to S_+)}{R(D^0\to S_+)}-\frac{\Gamma(D^0\to l^+)}{\Gamma(\bar{D}^0\to l^-)}\frac{R(D_+\to l^-)}{R(D_+\to l^+)}.
\end{split}
\label{eq.4.dtsignalall}
\end{equation}

Note that the asymmetries defined in Sec.~~\ref{sec:level4} are in terms of joint decay rates, while the asymmetries defined here are in terms of single particle decay rates, some of which are then obtained from joint decay events.

We can estimate those asymmetries in the SM. Using Eqs.~(\ref{eq.3.2}), (\ref{eq.1.3}), (\ref{eq.4.1}), (\ref{eq.4.2}), and (\ref{eq.3.singlerateint}),  we find
\begin{equation}
\begin{split}
&R(D^0\to S_{\pm}) \propto |A_{S_{\pm}}|^2\left(-\frac{2 \left(\left|\frac{p}{q}\right|^2+1\right) x^2+2 \left(\left|\frac{p}{q}\right|^2-1\right) y^2+1}{2\Gamma \left(4 x^2+1\right) \left(4 y^2-1\right)}\pm\right.\\
&\left.\frac{ \left|\frac{p}{q}\right| \left(\cos (2\phi) \left(4 x^2+1\right) y+\sin(2\phi) x \left(4 y^2-1\right)\right)}{\Gamma \left(4 x^2+1\right) \left(4 y^2-1\right)}\right),\\
\end{split}
\label{eq.4.singletransitionSM.1}
\end{equation}
\begin{equation}
\begin{split}
&R(\bar{D}^0\to S_{\pm}) \propto |A_{S_{\pm}}|^2\left(\frac{\left|\frac{p}{q}\right|^2 \left(-2 x^2+2 y^2-1\right)-2 \left(x^2+y^2\right)}{2\Gamma \left|\frac{p}{q}\right|^2 \left(4 x^2+1\right) \left(4 y^2-1\right)}\pm\right.\\
&\left. \frac{ \left(x \left(4 \cos(2\phi) x y-4 \sin(2\phi) y^2+\sin(2\phi)\right)+\cos(2\phi) y\right)}{\Gamma \left|\frac{p}{q}\right| \left(4 x^2+1\right) \left(4 y^2-1\right)}\right),\\
\end{split}
\label{eq.4.singletransitionSM.2}
\end{equation}
\begin{equation}
\begin{split}
&R(D_{\pm}\to l^+) \propto |A_l|^2\left(\frac{\left|\frac{p}{q}\right|^2 \left(-2 x^2+2 y^2-1\right)-2 \left(x^2+y^2\right)}{2\Gamma \left|\frac{p}{q}\right|^2 \left(4 x^2+1\right) \left(4 y^2-1\right)}\pm\right.\\
&\left. \frac{ \left(x \left(4 \cos(2\phi) x y-4 \sin(2\phi) y^2+\sin(2\phi)\right)+\cos(2\phi) y\right)}{\Gamma \left|\frac{p}{q}\right| \left(4 x^2+1\right) \left(4 y^2-1\right)}\right),\\
\end{split}
\label{eq.4.singletransitionSM.3}
\end{equation}
\begin{equation}
\begin{split}
&R(D_{\pm}\to l^-)\propto |A_l|^2\left(-\frac{2 \left(\left|\frac{p}{q}\right|^2+1\right) x^2+2 \left(\left|\frac{p}{q}\right|^2-1\right) y^2+1}{2\Gamma \left(4 x^2+1\right) \left(4 y^2-1\right)}\pm\right.\\
&\left. \frac{ \left|\frac{p}{q}\right| \left(\cos(2\phi) \left(4 x^2+1\right) y+\sin(2\phi) x \left(4 y^2-1\right)\right)}{\Gamma \left(4 x^2+1\right) \left(4 y^2-1\right)}\right).\\
\end{split}
\label{eq.4.singletransitionSM.4}
\end{equation}

We use the parameter values $x=0.0037$, $y=0.0066$, $\frac{q}{p}=0.91$, and $\phi=-4.7^o$, as given above.  As a result, the expected signal within the SM at the leading order can be written as
\begin{equation}
\begin{split}
&A_T^1\approx 8A_{\Gamma}+8x(A_Mx+2\sin(2\phi)y)\approx -1.5\times 10^{-4}
,\\
&A_T^2\approx 8A_{\Gamma}-8x(A_Mx+2\sin(2\phi)y)\approx 2.2\times 10^{-5}.
\end{split}
\label{eq.4.expectedres}
\end{equation}

The DT method using the entangled states has been used to measure $y_{\rm CP}$~\cite{BESCII}, which is of the order of about $10^{-3}$ to $10^{-2}$. We can conclude that, to observe  TV signals, which are  about $10^{-5}$ to $10^{-4}$, the event numbers should be   four orders greater than those for measuring  $y_{CP}$.

\section{\label{sec:level6}Simulation}

Through a Monte Carlo simulation~\cite{mcmethod}, we can estimate the significance of the expected time-dependent signal based on current experiments. The time-dependent signal in the $D^0-\bar{D}^0$ mixing is difficult to measure~\cite{PDG,BESCIIDetector} because the lifetimes of $D$ mesons are too short, thus requiring a very high resolution of the decay length. We have calculated above that the asymmetries in the $C=-1$ $D^0-\bar{D}^0$ state are very small. In this section, by using Monte Carlo simulation, we  analyze whether we are able to observe such signals or how far experimentally we are away from the required resolution.

Following the idea of Ref.~\cite{mcmethod}, we use $R_-(f_a,f_b,\Delta t)$ as the PDF to generate experimental events. For simplicity,  we only simulate the $D^0\to D_-$ and $D_-\to D^0$ transitions. We define $\tau \equiv \Gamma t$.

The PDF is affected by the mistakes in identifying the final states. In the case of $B$ mesons, only  the mistakes in the flavor identification were considered~\cite{mcmethod}. We assume this is also the case in $D$ mesons. The mistakes in identifying  a non-$CP$ eigenstate as $CP$ eigenstate  cancel each other between $S_{\pm}$ terms in the asymmetries.  Similarly, the mistakes in distinguishing  the semileptonic decays from background also cancel each other between $l^{\pm}$ terms. Moreover, the $CP$ violation in the decays of $K_S^0$ mesons~\cite{BESCII}, which is used in the $CP$ identification,  is known to be small; thus, the mistakes in distinguishing the two $CP$ eigenstates can be neglected. So, we only consider the mistakes in distinguishing the two flavor final states $l^+$ and $l^-$.

The PDF can be modified as~\cite{mcmethod}
\begin{equation}
\begin{split}
&\bar{R}_-(l^+,S_\pm,\Delta \tau)=(1-\omega _{l} )R_-(l^+,S_\pm, \Delta \tau)+\omega _{l }R_-(l^-,S_\pm, \Delta \tau),\\
&\bar{R}_- (l^-,S_\pm,\Delta \tau)=(1-\omega _{l })R_-(l^-,S_\pm, \Delta \tau)+\omega _{l }R_-(l^+,S_\pm, \Delta \tau),\\
\end{split}
\label{eq.5.1}
\end{equation}
where $\omega _{l }$ is the mistag rates in distinguishing $l^{\pm}$ final states.   We assume the confidence of identification of $l^{\pm}$ is similar to the case of $B$ mesons; hence, $\omega _{l } \approx 2.8\%$~\cite{Babar2009}.

The effect of $\Delta \tau$ resolution is complicated in the experiments~\cite{mcmethod,Babar2009,Babar2002}. We simply use a Gaussian function to include the effect of $\Delta \tau$ resolution,
\begin{equation}
\begin{split}
&h(\Delta \tau, \Delta \tau_{true}, \sigma _{\tau})=\frac{1}{\sqrt{2\pi} \sigma _{\tau}}\exp \left(-\frac{(\Delta \tau-\Delta \tau _{true})^2}{2\sigma _{\tau}^2}\right),
\end{split}
\label{eq.5.3}
\end{equation}
and the PDF can be modified as~\cite{mcmethod}
\begin{equation}
\begin{split}
&\mathcal{R}(l^{\pm},S_{\pm},\Delta \tau)\propto \bar{R}_- (l^{\pm},S_{\pm},\Delta \tau_{true}) H(\Delta \tau_{true})\otimes h(\Delta \tau, \Delta \tau_{true}, \sigma _{\tau})\\
&+\bar{R}_-(S_{\pm},l^{\pm},\Delta \tau_{true}) H(-\Delta \tau_{true})\otimes h(\Delta \tau, \Delta \tau_{true}, \sigma _{\tau}),
\end{split}
\label{eq.5.4}
\end{equation}
where $H(\Delta \tau)$ is Heaviside step function  and $\otimes$ denote convolution over $\Delta \tau_{true}$.

If  $\psi (3770)$ is at rest,  the proper time interval  $\Delta t$ of the decays of the two $D$ mesons is related with the momentum as~\cite{JDRyang}
\begin{equation}
\begin{split}
&\Delta t \approx \left(r_D-r_{\bar{D}}\right) \frac{m_D}{c|{\bf P}|},
\end{split}
\label{eq.5.5}
\end{equation}
where $r_D$ and $r_{\bar{D}}$ are decay lengths of $D^0$ and $\bar{D}^0$ mesons and ${\bf P}$ is the 3-momentum of $D^0$. The uncertainties mainly come from $r_D$ and $r_{\bar{D}}$. The average is $ \approx\;290\;{\rm \mu m}$, and one can  use  the rms of decay length in Belle, which is  $ <\;{\rm 100\;\mu m}$~\cite{JDRyang},  and then  $\sigma _{\tau}/\Delta \tau  \approx  100/290 \approx 34\%$.

We only generate the events with $\Delta \tau > 0$. The normalized PDF is
\begin{equation}
\bar{R}_{MC}(l^{\pm},S_{\pm},\Delta \tau)=\frac{1}{N}\mathcal{R}(l^{\pm},S_{\pm},\Delta \tau)H(\Delta \tau),
\label{eq.5.6}
\end{equation}
where
$N=\int _0^{+\infty}d(\Delta \tau)\mathcal{R}(l^{\pm},S_{\pm},\Delta \tau)$.

In Ref.~\cite{BESCII}, the number of double-tag events is about $5000$. Hence, we generate $5000$ events for both $D^0\to D_-$ and $D_-\to D^0$ using the PDF in Eq.~(\ref{eq.5.6}). With generated events, we are able to obtain the number of events $N_{MC}(f_a, f_b,\tau _0)$ in an interval $0 \sim \tau_0$. The numbers of events that we are interested in are   $  N_{MC}(S_+,l^+,\tau_0)$ and $  N_{MC}(l^-,S_-,\tau_0)$. We can also obtain the average decay time $\langle \Delta t \rangle _{MC}^{\pm}$ from generated events, where $\pm$ in the superscript represents the transition  with the $l^{\pm}$ final state.

\subsection{\label{sec6.1} Fitting joint decay rates }

Since we use the normalized PDF, we are not able to compare the time-independent joint decay rates of the conjugated transitions. So, we concentrate on comparing time-dependent joint decay rates.

Using Eq.~(\ref{eq.4.5}), we find that the normalized time-dependent joint decay rate of a $C=-1$ can be approximately  expressed as
\begin{equation}
\begin{split}
&r_-(l^-,S_-,\Delta t)=\frac{1}{n} e^{-\Gamma |\Delta t|}\left(2+b \Delta t+O(10^{-5}))\right),\\
&r_-(S_+,l^+,\Delta t)=\frac{1}{n} e^{-\Gamma |\Delta t|}\left(2+b \Delta t+O(10^{-5}))\right),\\
\end{split}
\label{eq.5.7}
\end{equation}
where $r_-(f_a,f_b,\Delta t)$ is defined in Eq.~(\ref{eq.4.8.add4}) and $b$ and $n$ satisfy
\begin{equation}
\begin{split}
&b\equiv 2\cos (2\phi)y\approx 2y_{\rm CP},\;\;n\equiv \frac{n_++n_-}{2},\\
&n_-\equiv\int _0^{+\infty}d(\Delta t)r_-(l^-,S_-,\Delta t),\;n_+\equiv\int _0^{+\infty}d(\Delta t)R_-(S_+,l^+,\Delta t),\\
&n_{\pm}=\frac{1}{\left|\frac{q}{p}\right|^{1\pm 1} \bar{x}\bar{y}}\left((\bar{x} \mp \bar{y})+\left|\frac{q}{p}\right|^2(\bar{x} \pm \bar{y})+2\left|\frac{q}{p}\right|\left(\cos (2\phi)y\bar{x} \pm \sin (2\phi)x \bar{y}\right)\right),\\
\end{split}
\label{eq.5.8}
\end{equation}
where $\bar{x}$ and $\bar{y}$ are defined in Eq.~(\ref{eq.4.18}). The number of events with $\Delta \tau < \tau_0$ can be obtained as
\begin{equation}
\begin{split}
&N_{SM}(f_a,f_b,\tau_0)= \mathcal{N}_f\int _0^{\tau _0}r_-(f_a,f_b,\Delta t),\\
\end{split}
\label{eq.5.9}
\end{equation}
where the subscript SM represents the expected result in the SM. $\mathcal{N}_f$ is the total number of events. With the definition $N^+_{SM}(\tau_0)\equiv N_{SM}(S_+,l^+,\Delta \tau_0)$ and $N^-_{SM}(\tau_0)\equiv N_{SM}(l^-,S_-,\Delta \tau_0)$, we find that, to the leading order,
\begin{equation}
\begin{split}
&N^+_{SM}(\tau_0)=N^-_{SM}(\tau_0)=\frac{1}{n} \mathcal{N}_f \left((2+b)(1-e^{-\tau_0})-b \tau_0 e^{-\tau_0}\right).\\
\end{split}
\label{eq.5.10}
\end{equation}

\begin{figure}
\resizebox{0.45\textwidth}{!}{%
\includegraphics{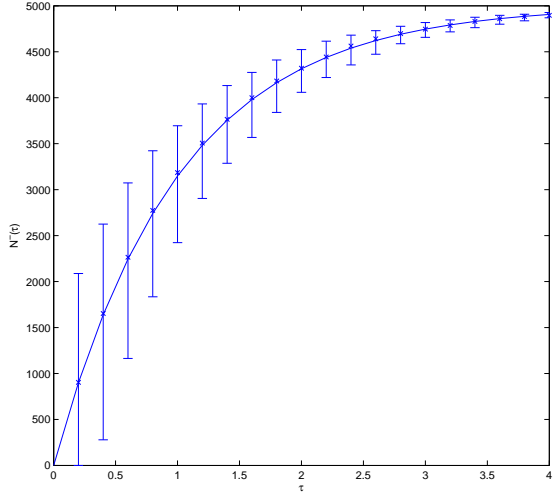}
}
\caption{\label{Fig:bmfit} One example of the fitting of $b^-$. The cross with the error bar is the generated $N_{MC}^-(\tau)$, where the error bars are generated because the $\delta \tau $ of the events is $34\%$. The solid line is the fitting $N_{SM}^-(\tau)$ using Eq.~(\ref{eq.5.10}). In this figure, the fitted result is $b^-=0.01312$.}
\end{figure}

\begin{figure}
\resizebox{0.45\textwidth}{!}{%
\includegraphics{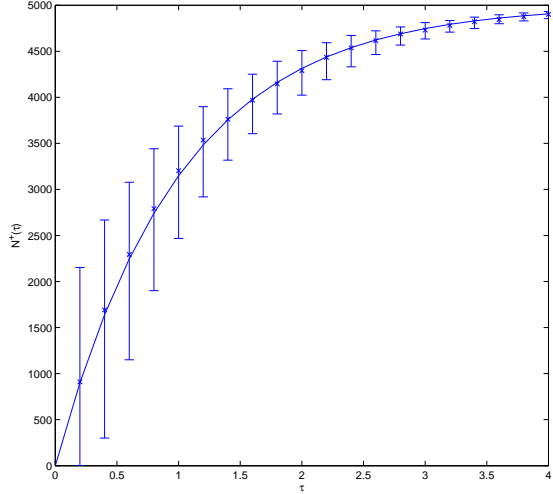}
}
\caption{\label{Fig:bpfit} One example of fitting of $b^+$. The cross with error bar is the generated $N_{MC}^+(\tau)$, where the error bars are generated because the standard deviation $\delta \tau $ of the events is $34\%$. The solid line is the fitted $N_{SM}^+(\tau)$ using Eq.~(\ref{eq.5.10}). In this figure, the fitted result is $b^+=0.01314$.}
\end{figure}

We can use Eq.~(\ref{eq.5.10}) to fit $N^{\pm}_{MC}(\tau_0)$, thereby determining the corresponding values of $b$, denoted as $b^{\pm}$, where the superscript corresponds to that of  $N^{\pm}_{MC}(\tau_0)$. If time reversal is conserved, one  has $b^+=b^-$. The difference between $b^{+}$  and $b^-$   can be identified as a signal of TV. Examples of the generated $N_{MC}^{\pm}(\tau _0)$ and  the fitting $N_{SM}^{\pm}(\tau _0)$ are shown in Figs.~\ref{Fig:bmfit} and \ref{Fig:bpfit}.

\begin{figure}
\resizebox{0.45\textwidth}{!}{%
\includegraphics{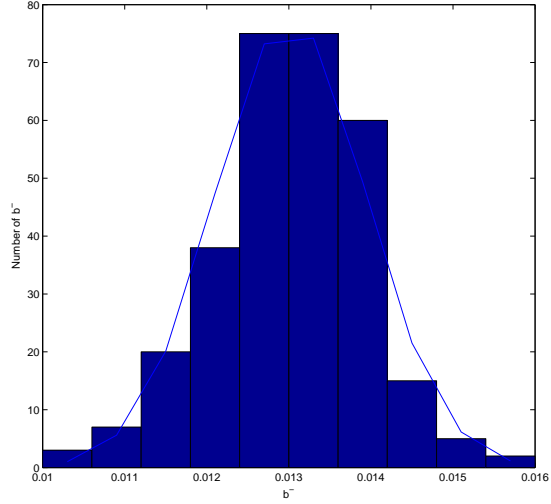}
}
\caption{\label{Fig:bmdist} The distribution of $b^-$ in 300 runs of the  simulation. The solid line is generated by the Gaussian distribution with the mean and the standard deviation given in Eq.~(\ref{eq.5.11}).}
\end{figure}

\begin{figure}
\resizebox{0.45\textwidth}{!}{%
\includegraphics{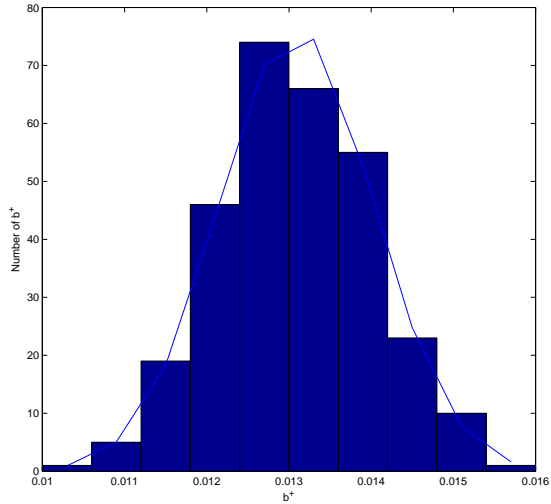}
}
\caption{\label{Fig:bpdist} The distribution of $b^+$ in 300 runs of simulation. The solid line is generated by the Gaussian distribution with the mean and the standard deviation given in Eq.~(\ref{eq.5.11}).}
\end{figure}

To estimate the uncertainty of $b^{\pm}$, we run such a simulation for $300$ times, and the distributions of $b^{\pm}$ are shown in Figs.~\ref{Fig:bmdist} and \ref{Fig:bpdist}, respectively,  and the results are
\begin{equation}
\begin{split}
&b^+=13.1\pm 0.9\times 10^{-3},\;\;\;b^-=13.0\pm 0.9 \times 10^{-3}.\\
\end{split}
\label{eq.5.11}
\end{equation}

Hence, it is difficult to observe  the TV  in time-dependent T asymmetry in the  $C=-1$   $D^0-\bar{D}^0$ state because $\Delta b  < \delta b^{\pm}$, where $\Delta b=|b^--b^+|$, $\delta b^{\pm}$ are the standard deviations of $b^{\pm}$.

We can also estimate how far we are from the observation of the signal. In the SM, we find
\begin{equation}
\begin{split}
&\Delta N_{SM}(\tau_0)\equiv N^-_{SM}(\tau_0)-N^+_{SM}(\tau_0)=s(1-e^{-\tau_0})-2u \tau_0 e^{-\tau_0}+O(10^{-6}),\\
\end{split}
\label{eq.5.12}
\end{equation}
where
\begin{equation}
\begin{split}
&s\equiv 4\Delta n+2u,\\
&u\equiv A_M\cos(2\phi)y-2\sin(2\phi)x+\frac{3}{8}A_M^3\cos(2\phi)y-\frac{1}{4}A_M^2\sin(2\phi)x,\\
&\Delta n \equiv \frac{n_+-n_-}{2}.\\
\end{split}
\label{eq.5.13}
\end{equation}
In the SM, we find $s=7.6\times 10^{-5}$ and $2u=-3.1\times 10^{-5}$; therefore,
\begin{equation}
\begin{split}
&b^{\pm}\approx b=0.013,\;\;\; \delta b^{\pm}< 10^{-4}.\\
\end{split}
\label{eq.5.14}
\end{equation}
Using Eqs.~(\ref{eq.5.11}) and (\ref{eq.5.14}), we find that with 5000 events the fitting  values of $b^{\pm}$ are very close to the expected values of $b^{\pm}$; however, the expected difference $\Delta b$ is too small to be observed. The accuracy of $b^{\pm}$ needs to be at least smaller than $10^{-4}$. So we can also conclude that, in  consistency  with  Sec.~\ref{sec:level4}, to observe the TV signal the number of events should be  at least four  orders of magnitude larger than the one in the current experiments, which is about 5000.

\subsection{\label{sec6.2}Average decay times}

In the above, we have used $\Delta \tau  \sim 1 $, such that $\Delta t \sim 1/\Gamma$. Here we verify this assumption, and use the difference between the average decay times   in the two conjugate processes as the evidence of TV. Each average decay time does not depend on fitting.

\begin{figure}
\resizebox{0.45\textwidth}{!}{%
\includegraphics{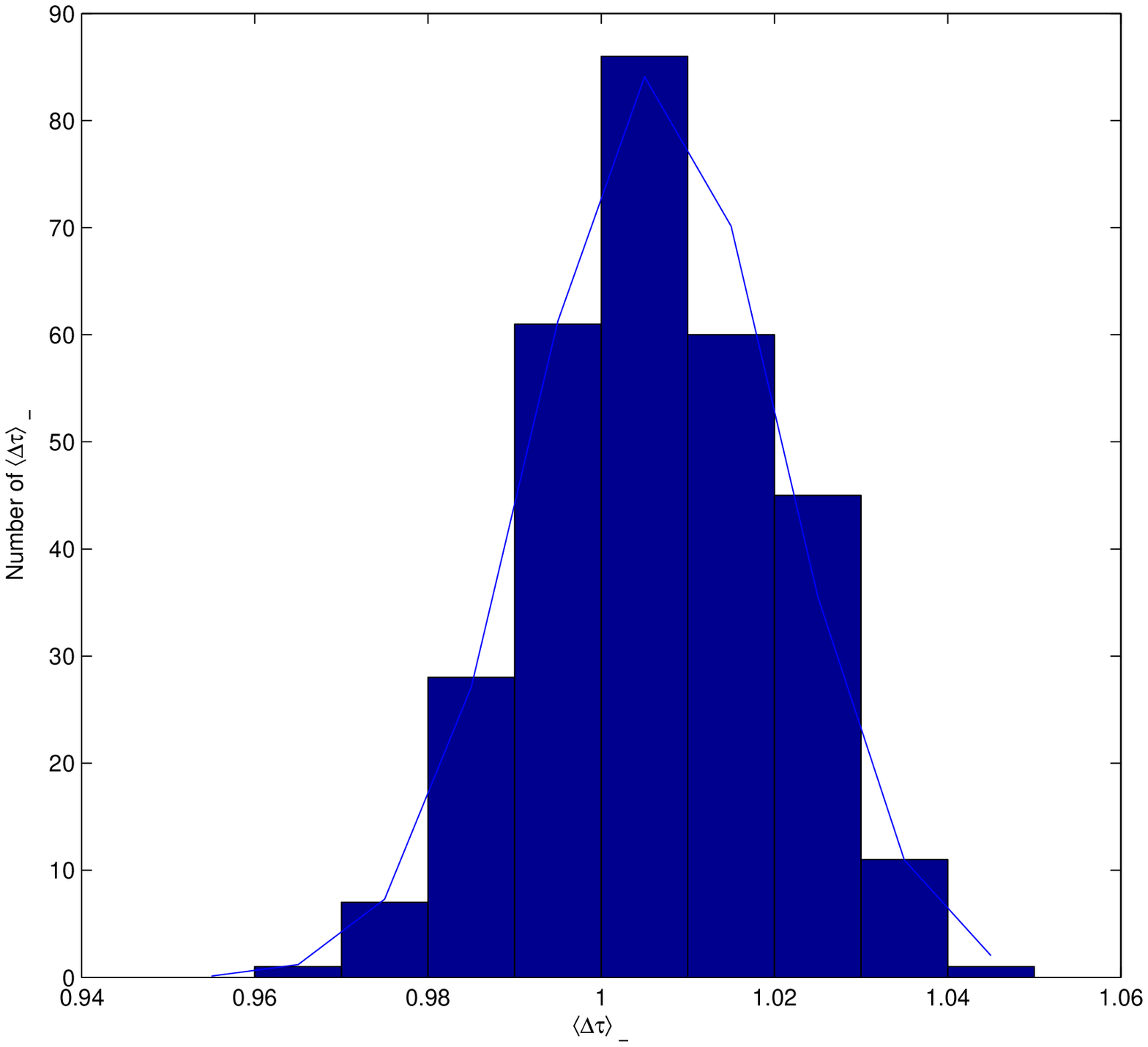}
}
\caption{\label{Fig:tmdist} The distribution of $\langle \Delta \tau \rangle _-$  in  300  runs of simulation. The solid line is generated by the Gaussian distribution with the mean and the standard deviation given  in Eq.~(\ref{eq.5.16}).}
\end{figure}

\begin{figure}
\resizebox{0.45\textwidth}{!}{%
\includegraphics{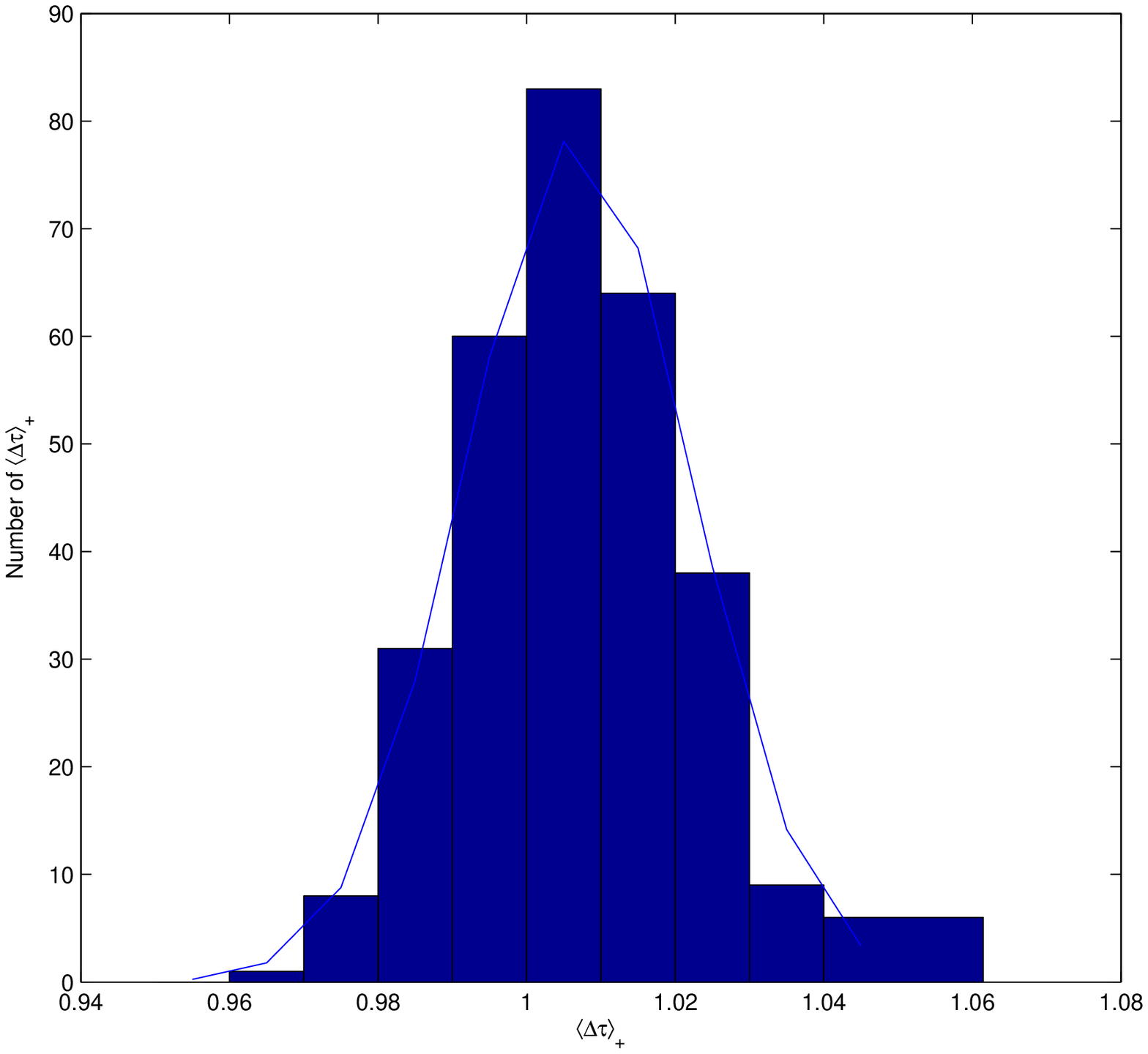}
}
\caption{\label{Fig:tpdist}  The distribution of $\langle \Delta \tau \rangle _+$  in  300  runs of simulation. The solid line is generated by the Gaussian distribution with the mean and the standard deviation given  in Eq.~(\ref{eq.5.16}). }
\end{figure}

In the SM, the average decay time can be obtained as
\begin{equation}
\begin{split}
&\langle \Delta \tau \rangle _- \equiv \int _0^{\infty}r_-(l^-,S_-,\Delta \tau)\Delta \tau d(\Delta \tau),\;
 \langle \Delta \tau \rangle _+ \equiv \int _0^{\infty}r_-(S_+,l^+,\Delta \tau)\Delta \tau d(\Delta \tau),\\
\end{split}
\label{eq.5.15}
\end{equation}
which  are obtained in  $300$ runs of the  simulation, as shown in Figs.~\ref{Fig:tmdist} and \ref{Fig:tpdist},  with the result
\begin{equation}
\begin{split}
&\langle \Delta \tau \rangle _{+,MC}=1.0068 \pm 0.0149, \;
 \langle \Delta \tau \rangle _{-,MC}=1.0063 \pm 0.0139,\\
\end{split}
\label{eq.5.16}
\end{equation}
Hence $|\langle \Delta \tau \rangle _{-,MC} -\langle \Delta \tau \rangle _{+,MC}|\ll \delta \langle \Delta \tau \rangle _{\pm,MC}$, where $\delta \langle \Delta \tau \rangle _{\pm,MC}$ is the standard deviation of $\langle \Delta \tau \rangle _{\pm,MC}$.  This suggests  the difficulty in observing the T-violating signal.

Let us estimate in the SM the accuracy needed  to observe the T-violating signal. We find
\begin{equation} \langle \Delta \tau \rangle _{\pm} =\frac{T_{\pm}}{n_{\pm}}, \end{equation}
where  $
T_{\pm}=\frac{1}{\left|\frac{q}{p}\right|^{1 \mp 1}}\left( \frac{\left|\frac{q}{p}\right|^2(1+y^2)+4\left|\frac{q}{p}
\right|\cos (2\phi)y}{\bar{y}^2}+ \frac{\left|\frac{q}{p}\right|^2(x^2-1)\pm 4\left|\frac{q}{p}\right|\sin(2\phi)x}{\bar{x}^2}\right)
+\left(\frac{1-x^2}{\bar{x}^2}+\frac{(1+y^2)}{\bar{y}^2}\right)$,
with $n_{\pm}$, and $\bar{x}$ and $\bar{y}$ are defined in Eq.~(\ref{eq.5.8}). The numerical results are
\begin{equation}
\begin{split}
&\langle \Delta \tau \rangle _{+,SM} \approx 1.0066,\;\;
 \langle \Delta \tau \rangle _{-,SM} \approx 1.0065.
\end{split}
\label{eq.5.18}
\end{equation}
To observe the T-violating  signal, the accuracy of measuring $\langle \Delta \tau \rangle$ should be about $10^{-5}$.

It should be noted that the number of events is an important factor that  greatly affects the accuracy.  We  have  run the simulation on $\langle \Delta \tau \rangle$ described above with different event numbers. The results are listed in Table~\ref{tab2}. To estimate the  standard deviation, each simulation with the same number of events is run 300 times. We find that the standard deviation is proportional to $1/\sqrt{N}$, where $N$ is the event number. According to the trend, if the event number is of  the order of $10^9\sim 10^{10}$, which can be expected in the super-tau-charm factory~\cite{supertaucharm}, the standard deviation reaches $10^{-5}$, which is the order of the magnitude of the lifetime difference between the T-conjugate processes, as predicted by the SM and $\omega$ effect,
\begin{equation}
\begin{split}
&\langle \Delta \tau \rangle _{+,SM} - \langle \Delta \tau \rangle _{-,SM} \approx 3.75\times 10^{-5},\\
&2.1\times 10^{-5}<\left.\left(\langle \Delta \tau \rangle _{+} - \langle \Delta \tau \rangle _{-}\right)\right|_{|\omega|=10^{-3}}<5.4\times 10^{-5}
\end{split}
\label{eq.5.19}
\end{equation}

Therefore, if the event number is of  the order of $10^9\sim 10^{10}$, which can be expected in the super-tau-charm factory,  then the TV signal can be observed, and the result can also set a bound on $|\omega|$ at about $10^{-3}$. That is to say, $|\omega| > 10^{-3}$ can be excluded if not observed.

\begin{table}
\caption{The result of the simulation with different event numbers. The standard deviation is obtained by running the simulation 300 times.}
\label{tab2}
\begin{tabular}{  c | c | c | c | c  }
\hline\noalign{\smallskip}
Number of events &  $10^4$ & $10^5$  & $10^6$  & $10^7$ \\
\noalign{\smallskip}\hline\noalign{\smallskip}
$\langle \Delta \tau \rangle _+-\langle \Delta \tau \rangle_-$ &  $(-5.3\pm 137)\times 10^{-4}$ & $(-0.95\pm 43)\times 10^{-4}$  & $(0.68\pm 14)\times 10^{-4}$ & $(2.9\pm 41)\times 10^{-5}$ \\
\hline
\noalign{\smallskip}\hline
\end{tabular}
\end{table}

\section{\label{sec:level7}Summary}

In this paper, we have studied TV in the $C=- 1$ entangled $D^0-\bar{D}^0$ systems, and various  T asymmetries are considered. We have proposed using the time-independent signals to study TV.

We calculated the time-dependent asymmetries of $C=-1$ system using joint decay rates, which are expected to be at the order of $10^{-5}$ in the SM. Using the joint decay rates, we also obtained the time-independent asymmetries,  which are also expected to be of  the order of $10^{-5}$ in the SM. We also studied  the contribution of the $\omega$ effect caused by a kind of CPTV, which changes the asymmetries  by as much as $20\%$  when $|\omega|\sim 10^{-4}$.

We also calculated T asymmetries defined for T-conjugate processes, the transitions  from $D^0$ to $D^-$ and vice versa, using the transition rates obtained from the event numbers in joint decays of entangled pairs. These time-independent T asymmetries are also of the order of $10^{-4}$  to $10^{-5}$.

We used  the Monte Carlo simulation   to estimate the time-dependent signals in the $C=-1$ entangled system by using the parameters in the current experimental situation. We estimate that if the event number reaches $10^9$  to   $10^{10}$ TV signals can be observed in the entangled $D^0-\bar{D}^0$ pairs and  the bound of $\omega \sim 10^{-3}$ can be reached.

In recent years, quantum entanglement has been found to be a resource of quantum information processing. Likewise, as exemplified by the present work, we may say that quantum entanglement is a resource of precision measurement in particle physics.

\acknowledgments

We thank Professor Haibo Li for useful discussions. This work is supported by National Natural Science Foundation of China (Grant No. 11574054).


\begin{thebibliography}{99}
\bibliographystyle{unsrt}

  \bibitem{CPreview}
  Y. Nir, Lectures given in the XXVII SLAC Summer Institute on Particle Physics, July 7 - 16, 1999;
  J. Bernab$\rm{\acute{e}}$u, J. Phys. Conf. Ser. {\bf 631}, 012015 (2015);
  A. Bevan, J. Phys. Conf. Ser. {\bf 631}, 012003 (2015);
  T. Gershon and V. V. Gligorov, Rep. Prog. Phys. {\bf 80}, 046201 (2017);
  F. C. Porter Prog. Part. Nucl. Phys. {\bf 91}, 101 (2016).

  \bibitem{PDG}
  K. A. Olive \textit{et al}. (Particle Data Group), Chin. Phys. C {\bf 38}, 090001 (2014) and 2015 update.

  \bibitem{NP}
  A. Lenz, Proceedings of the 8th International Workshop on the CKM Unitarity Traingle, 8-12 September 2014, Vienna, Austria; 
  M. Blanke, Nuovo Cimento C {\bf 39}, 329 (2017). 

  \bibitem{CPTV}
  J. Bernab$\rm{\acute{e}}$u, J. Ellis, N. E. Mavromatos, D. V. Nanopoulos, J. Papavassiliou, arXiv:hep-ph/0607322.

  \bibitem{omega} 
  J. Bernab$\rm{\acute{e}}$u, N. E. Mavromatos, and J. Papavassiliou, Phys. Rev. Lett. {\bf 92}, 131601 (2004);
  J. Bernab$\rm{\acute{e}}$u \textit{et al}. Nucl. Phys. B{\bf 744}, 180 (2006);
  J. Bernab$\rm{\acute{e}}$u \textit{et al}, arXiv:hep-ph/0607322.

  \bibitem{CPT}
  G. L$\rm{\ddot{u}}$ders, Dan. Mat. Phys. Medd. {\bf 28}, 5 (1954).

  \bibitem{original}
  M. C. Ba\~{n}uls and J. Bernab\'{e}u, Phys. Lett. B {\bf 464}, 117 (1999);
  M. C. Ba\~{n}uls and J. Bernab\'{e}u, Nucl.  Phys.  B{\bf 590}, 19 (2000);
  L. Wolfenstein, Int. J. Mod. Phys. E {\bf 8}, 501 (1999).

  \bibitem{Treview1}
  J. Bernab$\rm{\acute{e}}$u and F. Martinez-Vidal, Rev. Mod. Phys. {\bf 87}, 165 (2015).

  \bibitem{kextension}
  J. Bernab$\rm{\acute{e}}$u, A. Di Domenico, and P. Villanueva-Perez, Nucl. Phys. B{\bf 868}, 102 (2013).

  \bibitem{Treview2} 
  E. M. Henley, Int. J. Mod. Phys. E {\bf 22}, 1330010 (2013).

  \bibitem{TV}
  P. del Amo Sanchez \textit{et al}. (\textit{BABAR} Collaboration), Phys. Rev. D {\bf 81}, 111103 (2010);

  \bibitem{TVK}
  A. Angelopoulos \textit{et al}. (CPLEAR Collaboration), Phys. Lett. B {\bf 444} 43 (1998).

  \bibitem{TVB}
  J. P. Lees \textit{et al}. (\textit{BABAR} Collaboration), Phys. Rev. Lett. {\bf 109}, 211801 (2012). 

  \bibitem{TVBDiscuss}
  E. Applebaum \textit{et al}. 	Phys. Rev. D {\bf 89}, 075011 (2014).

  \bibitem{BdSystem} 
  J. Bernab$\rm{\acute{e}}$u, F. J. Botella, and M. Nebot, J. High Energy Phys. 06 (2016) 100. 


  \bibitem{k2}
  J. Bernab$\rm{\acute{e}}$u, A. Di Domenico, and P. Villanueva-Perez, J. High Energy Phys. 10 (2015) 139. 

  \bibitem{BESCII}
  M. Ablikim \textit{et al}. (BESIII Collaboration), Phys. Lett. B {\bf 744}, 339 (2015). 

  \bibitem{JDRxzz1}
  Z.-Z. Xing, Phys. Rev. D {\bf 55}, 196 (1997).

  \bibitem{JDRxzz2}
  Z.-Z. Xing, Phys. Lett. B {\bf 372}, 317 (1996). 

  \bibitem{mcmethod} 
  J. Bernab$\rm{\acute{e}}$u, F. Martinez-Vidal, and P. Villanueva-Perez, J. High Energy Phys. 08 (2012) 064. 

  \bibitem{Af}
  D. M. Asner and W. M. Sun, Phys. Rev. D {\bf 73}, 034024 (2006). 

  \bibitem{JDRreview}
  U. Nierste, Lectures at Helmholtz International Summer School "Heavy quark physics", Dubna, Russia, August 11-21, 2008. 

  \bibitem{cplus} 
  Z.-J. Huang and Y. Shi, Phys. Rev. D {\bf 89}, 016018 (2014). 

  \bibitem{JDRyang}
  H.-B. Li and M.-Z. Yang, Phys. Rev. D {\bf 74}, 094016 (2006). 

  \bibitem{HFAG} 
  Y. Amhis, \textit{et al}. (Heavy Flavor Averaging Group Collaboration), arXiv:1412.7515 and May 2015 update.

  \bibitem{JDRshi1}
  Y. Shi, Eur .Phys. J. C {\bf 73}, 2506 (2013). 

  \bibitem{entangled} 
  H. J. Lipkin, Phys. Lett. B {\bf 219}, 474 (1989). 
  
  \bibitem{DT}
  J. Adler \textit{et al}. (MARK-III Collaboration), Phys. Rev. Lett. {\bf 60}, 89 (1988).

  \bibitem{DT2}
  R. M. Baltrusaitis \textit{et al}. (MARK-III Collaboration), Phys. Rev. Lett. {\bf 56}, 2140 (1986).

  \bibitem{CPVSummery} 
  I. I. Bigi, A. Paul, and S. Recksiegel, J. High Energy Phys. 06 (2011) 089. 

  \bibitem{YcpDefine}
  S. Bergmann \textit{et al}. Phys. Lett. B {\bf 486}, 418 (2000); 
  M. Staric (Belle Collaboration), arXiv:1212.3478.
  
  \bibitem{Babar2009}  
  B. Aubert \textit{et al}. (\textit{BABAR} Collaboration), Phys. Rev. D {\bf 79}, 072009 (2009). 

  \bibitem{Babar2007}
  B. Aubert \textit{et al}. (\textit{BABAR} Collaboration), Phys. Rev. Lett. {\bf 99}, 171803 (2007) 

  \bibitem{supertaucharm}
  S. Eidelman, Nucl. Part. Phys. Proc. {\bf 260}, 238 (2015). 

  \bibitem{omegabd}
  E. Alvarez, J. Bernab$\rm{\acute{e}}$u, and M. Nebot, J. High Energy Phys. 11 (2006) 087. 

  \bibitem{BESCIIDetector}
  M. Ablikim \textit{et al}. (BESIII Collaboration), Nucl. Instrum. Methods Phys. Res., Sect. A {\bf 614}, 345 (2010). 

  \bibitem{Babar2002}
  B. Aubert \textit{et al}. (\textit{BABAR} Collaboration), Phys. Rev. D {\bf 66}, 032003 (2002); 
  J. Bernab$\rm{\acute{e}}$u , F. J. Botella, N. E. Mavromatos, M. Nebot, Eur. Phys. J. C {\b f77}, 865 (2017).



\end{thebibliography}
\end{document}